\documentclass[pra,twocolumn,aps,amssymb,showpacs,superscriptaddress]{revtex4-1}

\usepackage{epsfig}
\usepackage{graphicx}
\usepackage{amsmath}
\usepackage{amssymb}
\usepackage{enumerate}
\usepackage{hyperref}

\usepackage{color}
\definecolor{rosso}{rgb}{1,0,0}
\definecolor{verde}{rgb}{0,1,0}
\definecolor{blue}{rgb}{0,0,1}
\definecolor{amber}{rgb}{1.0, 0.75, 0.0}
\definecolor{amber(sae/ece)}{rgb}{1.0, 0.49, 0.0}
\definecolor{verdescuro}{rgb}{0,0.5,0.5}
\definecolor{rossoscuro}{rgb}{0.7,0.3,0}
\definecolor{bluscuro}{rgb}{0.3,0,0.7}
\definecolor{magenta}{rgb}{1,0,1}

\begin{document}

\title{Gap equation with pairing correlations beyond mean field and \\ its equivalence to a Hugenholtz-Pines condition for fermion pairs}

\author{L. Pisani}
\affiliation{School of Science and Technology, Physics Division, Universit\`{a} di Camerino, 62032 Camerino (MC), Italy}
\author{P. Pieri}
\email{pierbiagio.pieri@unicam.it}
\affiliation{School of Science and Technology, Physics Division, Universit\`{a} di Camerino, 62032 Camerino (MC), Italy}
\affiliation{INFN, Sezione di Perugia, 06123 Perugia (PG), Italy}
\author{G. Calvanese Strinati}
\email{giancarlo.strinati@unicam.it}
\affiliation{School of Science and Technology, Physics Division, Universit\`{a} di Camerino, 62032 Camerino (MC), Italy}
\affiliation{INFN, Sezione di Perugia, 06123 Perugia (PG), Italy}
\affiliation{CNR-INO, Istituto Nazionale di Ottica, Sede di Firenze, 50125 (FI), Italy}


\begin{abstract}
The equation for the gap parameter represents the main equation of the pairing theory of superconductivity.
Although it is formally defined through a single-particle property, physically it reflects the pairing correlations between opposite-spin fermions.
Here, we exploit this physical connection and cast the gap equation in an alternative form which explicitly highlights these two-particle correlations, by showing that it is equivalent
to a Hugenholtz-Pines condition for fermion pairs.
At a formal level, a direct connection is established in this way between the treatment of the condensate fraction in condensate systems of fermions and bosons.
At a practical level, the use of this alternative form of the gap equation is expected to make easier the inclusion of pairing fluctuations beyond mean field.
As a proof-of-concept of the new method, we apply the modified form of the gap equation to the long-pending problem about the inclusion of the Gorkov-Melik-Barkhudarov correction across the whole BCS-BEC crossover, from the BCS limit of strongly overlapping Cooper pairs to the BEC limit of dilute composite bosons, and for all temperatures in the superfluid phase.
Our numerical calculations yield excellent agreement with the recently determined experimental values of the gap parameter for an ultra-cold Fermi gas in the intermediate regime between BCS and BEC, 
as well as with the available quantum Monte Carlo data in the same regime.
\end{abstract}

\pacs{74.20.Fg,03.75.Ss,05.30.Jp} 

\maketitle

\section{Introduction} 
\label{sec:introduction}

As Lev Gor'kov first realized \cite{Gorkov-1958}, in the ultimate analysis the BCS theory of superconductivity \cite{BCS-1957,Schrieffer-1964} rests on the assumption that the average value $\langle \psi_{\uparrow}(\mathbf{r}) \psi_{\downarrow}(\mathbf{r}) \rangle$ is non-vanishing, where $\psi_{\sigma}(\mathbf{r})$ is the fermion field operator with spin $\sigma=(\uparrow,\downarrow)$ at spatial position $\mathbf{r}$.
This basic idea was sufficient to Gor'kov for formulating the BCS theory in terms of single-particle fermionic propagators (or many-body Green's functions), thereby adapting the presence of Cooper pairs between opposite-spin fermions \cite{Cooper-1956} to the apparatus of quantum field theory.
Physically, the non-vanishing of $\langle \psi_{\uparrow}(\mathbf{r}) \psi_{\downarrow}(\mathbf{r}) \rangle$ entails a sort of Bose condensation of pairs below a certain critical temperature $T_{c}$.
It is thus clear that an equation determining $\langle \psi_{\uparrow}(\mathbf{r}) \psi_{\downarrow}(\mathbf{r}) \rangle$ (or, better, a physical quantity directly related to it) plays a key role in the theory.
It turns out that this quantity is the so-called BCS gap $\Delta$, which is the product of $\langle \psi_{\uparrow}(\mathbf{r}) \psi_{\downarrow}(\mathbf{r}) \rangle$ with the strength $v_{0}$ of the inter-particle attraction between opposite-spin fermions, when this is taken of the contact type for convenience.

The Gor'kov theory \cite{Gorkov-1958} was framed at the mean-field level like the original BCS theory itself \cite{BCS-1957,Schrieffer-1964}, whereby all Cooper pairs ``on the average'' are dealt with on equal footing.
Quantum and thermal fluctuations over and above mean field, however, act on the individual partners of a fermion pair and somewhat disrupt their pairing, thereby resulting in a decrease of the value of $\Delta$ (as well as of $T_{c}$).
Accordingly, whenever these \emph{pairing fluctuations} become important, it is necessary to include their effect in the gap equation that determines $\Delta$.

A good physical parameter to gauge to what extent an attractive inter-particle coupling affects the Fermi gas is the ratio of the Cooper pair size $\xi_{\mathrm{pair}}$ to the inter-particle distance (given in terms of
the inverse of the Fermi wave vector $k_{F} = (3 \pi^{2} n)^{1/3}$ where $n$ is the density).
Here, weak (strong) coupling is identified by $k_{F} \xi_{\mathrm{pair}}$ being much larger (smaller) than unity.
These two situations correspond to the BCS and BEC limits of the BCS-BEC crossover, with strongly overlapping Cooper pairs and dilute composite bosons present in the two regimes, respectively.
Under these circumstances, fluctuations that act to disrupt pairing are expected to affect the value of $\Delta$ more significantly in the weak- (BCS) than in the strong- (BEC) coupling regime.
In this respect, the result by Gor'kov and Melik-Barkhudarov (GMB) \cite{GMB-1961}, who found for $\Delta$ a reduction by a factor $2.2$ with respect to its mean-field value at zero temperature in the (extreme) BCS limit $k_{F} \xi_{\mathrm{pair}} \ll 1$, is particularly significant.
This result, however, was obtained in Ref.~\cite{GMB-1961} not by solving an appropriate gap equation with a beyond-mean-field contribution, but rather by looking at the instability of the vertex function which is at the core of the two-particle propagator \cite{AGD-1963}.
As a consequence, it appears difficult to extend the original GMB analysis for $\Delta$ to the whole BCS-BEC crossover and for all temperature below $T_{c}$.

In this context, the interest in the BCS-BEC crossover is motivated from two sides.
On the one hand, numerical calculations based on approximate treatments of many-body diagrammatic methods can be tested against the analytic results that can be obtained in \emph{both} the opposite weak- (BCS) and strong- (BEC) limits, where the physical soundness of the obtained results can be controlled.
On the other hand, a stringent comparison is possible with the experimental results obtained with ultra-cold Fermi gases, in particular, in the intermediate-coupling (unitary) regime for which no analytic result is available.
This topic is also of concern in nuclear physics, where in low-density neutron matter the unitary regime can be approached from the weak-coupling (BCS) limit \cite{Carlson-2010,PhysRep-2018}.

The main purpose of this paper is to set up a modified form of the gap equation which explicitly highlights two-particle (pairing) correlations, in such a way that pairing fluctuation corrections beyond mean field can be most readily, not only introduced at a formal level, but also calculated numerically in relevant cases of interest.
We also show that this modified form of the gap equation is the equivalent for fermion pairs of the Hugenholtz-Pines condition for point-like bosons \cite{HP-1959}, which is also the main reason why 
it is ideally suited to span the BCS-BEC crossover from Cooper pairs to composite bosons.

The equivalence between the modified form of the gap equation and the Hugenholtz-Pines condition for fermion pairs is proved to be quite general, to the extent that it holds for any self-consistent 
(or, better, conserving \cite{Baym-1962}) approximation chosen to describe the underlying fermionic system, with the only provision that the inter-particle potential is of the contact type.
On physical grounds, with this choice one can focus the efforts directly on addressing the effects of pairing fluctuations in the superfluid phase, leaving aside the (possibly irrelevant) complications introduced by more complex forms of the interaction potential.

We remark that, for the way it is formulated, the present approach differs from the more traditional ones, which aim at formally setting up an integral equation for the ``anomalous'' single-particle self-energy in the broken-symmetry phase below $T_{c}$ \cite{Nozieres-1964}, while keeping an arbitrary form of the inter-particle interaction.
Here, by limiting ourselves to the use of a contact interaction, we will be able to somewhat reduce the complexity of the ensuing mathematical problem, in a way that will make it easier to concentrate directly on the effects of pairing fluctuations over and above mean field.
For this reason, the present treatment is also amenable to a direct extension to the normal phase above $T_{c}$, a result which is not possible to achieve for approaches that concentrate instead on the anomalous single-particle self-energy below $T_{c}$.

To provide a proof-of-concept of the improvements that the new method can introduce in practice, for determining the superfluid gap parameter when beyond-mean-field corrections are required, we will 
specifically consider an application of the method to the long-standing problem of extending the original GMB many-body diagrammatic analysis of Ref.~\cite{GMB-1961}, which was limited to the extreme BCS regime at zero temperature, to the whole BCS-BEC crossover and for all temperatures below $T_{c}$.
(It will turn out that, besides the GMB correction, an additional (Popov) diagrammatic correction \cite{Pieri-2005} will be required for a correct recovering of the value of the gap in the BCS limit.)

At a practical level, the choice of the GMB problem, as a test of the modified form of the gap equation transformed into a Hugenholtz-Pines condition for fermion pairs, is suggested by the fact that an accurate analysis of the corresponding problem in the normal phase above $T_{c}$ has recently been made available in Ref.~\cite{Pisani-2018}.
In that context, it was established that a proper (although numerically nontrivial) inclusion of the wave-vector and frequency dependence of the pair propagators, that occur in the diagrammatic expression of the GMB
correction, is essential to get meaningful results away from the (extreme) BCS limit to which the original GMB analysis was restricted.
We shall consistently verify that this effect is as well important in the superfluid phase below $T_{c}$.
In this way, we shall take advantage of the experience developed in Ref.~\cite{Pisani-2018} also at the computational level.

Besides the two main achievements of this paper, namely, having interpreted the gap equation as a Hugenholtz-Pines conditions for fermion pairs and having implemented it as a proof-of-concept for the non-trivial problem of the GMB correction over the whole superfluid sector of the coupling-vs-temperature phase diagram of the BCS-BEC crossover, a number of additional interesting features have also emerged along the way from our approach.
They include the numerical implementation of the Popov correction introduced in Ref.~\cite{Pieri-2005} and the identification of additional contributions to the scattering length $a_{B}$ for composite bosons in the BEC limit (over and above the Born value $a_{B}=2a_{F}$ obtained at the mean-field level), which affect both the condensate and non-condensate densities.

It should be mentioned that a few works have already extended the original GMB work of Ref.~\cite{GMB-1961} below $T_{c}$ in different directions. 
Specifically, the inclusion of screening effects in the gap equation was considered in Refs.~\cite{Rodero-1992,Kim-2009} for lattice models and in Refs.~\cite{Schulze-2001,Cao-2006} for neutron and nuclear matter. 
In these works, however, the problem of the extension of the GMB corrections to the whole BCS-BEC crossover was not considered.  
More recently, this extension was addressed in Ref.~\cite{Floerchinger-2010} within the functional-renormalization-group formalism, which is, however, completely different from the many-body diagrammatic approach here considered. 
Finally, the inclusion of the GMB correction throughout the BCS-BEC crossover with a many-body diagrammatic method was recently considered in Ref.~\cite{Chen-2016}. 
In this work, however, the wave-vector and frequency dependence of the pair propagator was not taken into account, an approximation that can be justified in practice (as we shall see) only in the extreme weak-coupling limit. 

The plan of the paper is as follows. 
Section~\ref{sec:HP-condition} provides a formal proof of the equivalence between the gap equation (in its appropriately modified version) and the Hugenholtz-Pines condition for fermion pairs with the use of diagrammatic methods.
This equivalence, which is shown to hold for any fermionic conserving (or, at least, self-consistent) approximation, makes it easier to include the effects of pairing fluctuations beyond mean field on the gap itself. 
Section~\ref{sec:G-MB-BCS-BEC} implements this formal equivalence in a practical context, by addressing the long-standing (and still pending) problem about the inclusion of the GMB contribution to the gap across the BCS-BEC crossover.
Section~\ref{sec:numerical-results} describes the strategies we have adopted to solve numerically the modified form of the gap equation.
It also presents our results for the temperature and coupling dependence of the gap parameter, throughout the BCS-BEC crossover and for all temperatures in the superfluid phase below the critical temperature $T_{c}$.
In this context, the favourable comparison between our results and the available experimental and quantum Monte Carlo (QMC) data will be emphasized as an indirect check on the validity of our diagrammatic approach.
Section~\ref{sec:conclusions} gives our conclusions and discusses the dichotomy between the thermodynamic (order parameter) and dynamic gap.
More technical details are given in the Appendices.
Appendix~\ref{sec:appendix-A} summarizes the main features of the $t$-matrix approximation in the broken-symmetry phase.
Appendix~\ref{sec:appendix-B} considers in detail the BEC limit of the Popov and GMB bosonic-like self-energies in the broken-symmetry phase, and shows how they are related to diagrammatic processes
associated with the scattering length for composite bosons.
Appendix~\ref{sec:appendix-C} describes a number of manipulations on the expressions to be calculated numerically, aiming at bringing them to a form as close as possible to the corresponding expressions valid in the normal phase above $T_{c}$.

In the following, only balanced populations between spin-up and spin-down fermions will explicitly be considered (even though the present treatment of the modified form of the gap equation could be extended as well to population- and mass-imbalanced fermions).
In addition, the reduced Planck constant $\hbar$ and the Boltzmann constant $k_{B}$ will everywhere be set equal to unity.

\section{Gap equation as a Hugenholtz-Pines condition for fermion pairs} 
\label{sec:HP-condition}
\vspace{-0.3cm}

In this Section, we prove the equivalence between the gap equation to determine the gap parameter in its suitably modified form and the Hugenholtz-Pines condition for fermion pairs. 
This equivalence is proved at a formal level, with the use of many-body diagrammatic techniques.
The emphasis on this equivalence is motivated by the fact that it makes more direct (and possibly easier at a practical level, as we shall see in Sections~\ref{sec:G-MB-BCS-BEC} and 
\ref{sec:numerical-results}) the introduction of pairing-fluctuation corrections over and above the standard mean-field level. 

The system we consider is a Fermi gas with inter-particle interaction $v_{0} \delta(\mathbf{r}-\mathbf{r'})$ of the contact type ($v_{0}<0$), which acts between opposite-spin fermions.
This singular potential has to be handled through a suitable regularization procedure, which can be expressed in terms of the scattering length $a_{F}$ of the two-fermion problem, in the form \cite{Sa_de_Melo-1993}:
\begin{equation}
\frac{m}{4\pi a_{F}} =  \frac{1}{v_{0}} + \int_{|\mathbf{k}| \le k_{0}} \! \frac{d\mathbf{k}}{(2 \pi)^3} \, \frac{m}{\mathbf{k}^2} 
\label{regularization}
\end{equation}
\noindent
where $k_{0}$ is an ultraviolet cutoff on the magnitude of the wave vector $\mathbf{k}$.
The limits $v_{0} \rightarrow 0^{-}$ and $k_{0} \rightarrow \infty$ are then simultaneously considered in order to keep $a_{F}$ at the desired value.

There are two main reasons to consider this system.
On the theoretical side, access to the regularization procedure (\ref{regularization}) considerably simplifies the handling of the diagrammatic structure, by getting rid from the outset of whole classes of  diagrammatic structures which do not survive the limit $v_{0} \rightarrow 0^{-}$ of the inter-particle interaction.
On the experimental side, this kind of system is well represented by a gas of ultra-cold Fermi atoms which are routinely utilised in experiments, in terms of which it is possible to span the BCS-BEC crossover.
In this respect, having separate access to the two opposite BCS and BEC regimes is also of theoretical importance, because in these limits distinct analytic results can be obtained.

The BCS-BEC crossover of interest is driven by the dimensionless coupling parameter $(k_{F} a_{F})^{-1}$. 
This parameter ranges from $(k_{F}\, a_{F})^{-1} \lesssim -1$ in the weak-coupling (BCS) regime when $a_{F} < 0$, to $(k_{F}\, a_{F})^{-1} \gtrsim +1$ in the strong-coupling (BEC) regime when $a_{F} > 0$, across the unitary limit when $|a_{F}|$ diverges. 

\vspace{0.05cm}
\begin{center}
{\bf A. Modified form of the gap equation}
\end{center}
\vspace{-0.2cm}

In the broken-symmetry phase we are interested in, it is convenient to express the field operators (on which the many-body diagrammatic structure is built) in the Nambu representation \cite{Nambu-1960}:
\begin{equation}
\Psi_{1}(\mathbf{r}) = \psi_{\uparrow}(\mathbf{r}) \,\,\, , \,\,\, \Psi_{2}(\mathbf{r}) = \psi_{\downarrow}^{\dagger}(\mathbf{r}) \, .
\label{Nambu-notation}
\end{equation}
\noindent
When this notation is translated into Fourier space, the Gor'kov equations \cite{Gorkov-1958} for the ``normal'' ($\mathcal{G}_{11}$) and ``anomalous'' ($\mathcal{G}_{12}$) single-particle fermionic propagators in the broken-symmetry phase read:
\begin{eqnarray}
& &  \left( 
\begin{array}{cc}
i \omega_{n} - \xi_{\mathbf{k}} - \Sigma_{11}(k)  &  - \Sigma_{12}(k)  \\
- \Sigma_{21}(k)  & i \omega_{n} + \xi_{\mathbf{k}} - \Sigma_{22}(k)
\end{array} 
\right)
\nonumber \\
& \times &
\left( 
\begin{array}{cc}
\mathcal{G}_{11}(k) & \mathcal{G}_{12}(k)  \\
\mathcal{G}_{21}(k) & \mathcal{G}_{22}(k) 
\end{array} 
\right) 
= 
\left( 
\begin{array}{cc}
1 & 0 \\
0 & 1
\end{array} 
\right)                                           
\label{Gorkov-Nambu-equations} 
\end{eqnarray}
\noindent
with the four-vector notation $k=(\mathbf{k},\omega_{n})$ where $\omega_{n}=(2 n + 1) \pi T$ ($n$ integer) is a fermionic Matsubara frequency  \cite{FW-1971}.
In this expression, $\xi_{\mathbf{k}} = \mathbf{k}^{2}/(2m) - \mu$ where $m$ is the fermion mass and $\mu$ the chemical potential.
In addition, the self-energies $\Sigma_{ij}(k)$ satisfy the properties $\Sigma_{22}(k)=-\Sigma_{11}(-k)$ and $\Sigma_{21}(k)=\Sigma_{12}(k)$
(assuming $\Delta$ - defined by Eq.~(\ref{gap-equation}) below - real without loss of generality).

Within the Nambu notation, the \emph{gap equation} for the gap parameter $\Delta$ reads:
\begin{equation}
\Delta \equiv v_{0} \, \langle \psi_{\uparrow}(\mathbf{r}) \psi_{\downarrow}(\mathbf{r}) \rangle = - v_{0} \sum_{k}\, \mathcal{G}_{12}(k) \, .
\label{gap-equation}
\end{equation}
\noindent
Here and in the following, we adopt for convenience the short-hand notation:
\begin{equation}
\sum_{k} = \int \! \frac{d\mathbf{k}}{(2 \pi)^{3}} \, T \, \sum_{\omega_{n}} \, .
\label{convention-four-integral}
\end{equation}

The gap equation (\ref{gap-equation}) is apparently expressed in terms of single-particle properties. 
However, it can be cast in a different form, from which two-particle properties (and thus pairing) readily appear. 
To this end, we first solve for $\mathcal{G}_{12}$ in Eq.~(\ref{Gorkov-Nambu-equations}), whereby the following \emph{identity} results
\begin{equation}
\mathcal{G}_{12}(k) = \mathcal{G}_{11}(k) \Sigma_{12}(k) \mathcal{G}_{22}(k) - \mathcal{G}_{12}(k) \Sigma_{21}(k) \mathcal{G}_{12}(k)
\label{general-identity}
\end{equation}
\noindent
which involves all the single-particle propagators $\mathcal{G}_{ij}$.
Then we combine Eq.~(\ref{general-identity}) with Eq.~(\ref{gap-equation}) and arrive at the following \emph{modified form of the gap equation}:
\begin{equation}
- \frac{\Delta}{v_{0}} = \sum_{k} \left[ \mathcal{G}_{11}(k) \Sigma_{12}(k) \mathcal{G}_{22}(k) - \mathcal{G}_{12}(k) \Sigma_{21}(k) \mathcal{G}_{12}(k) \right] \, .
\label{modified-gap-equation}
\end{equation}
\begin{figure}[t]
\begin{center}
\includegraphics[width=8.6cm,angle=0]{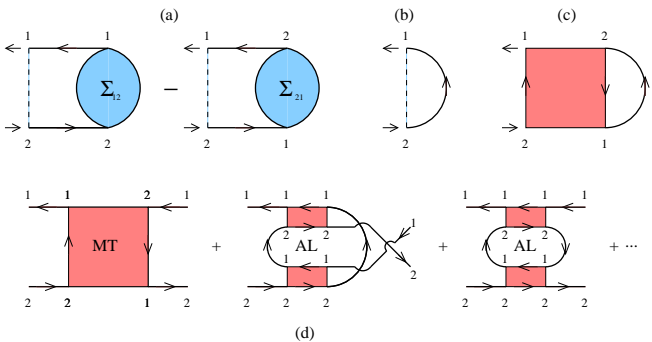}
\caption{(Color online) Diagrammatic representation of the: 
                                    (a) right-hand side of the modified form of the gap equation (\ref{modified-gap-equation}) multiplied by $v_{0}$;
                                    (b) Fock-like term for the anomalous fermionic self-energy $\Sigma_{12}$;
                                    (c) conserving $t$-matrix approximation for $\Sigma_{12}$;
                                    (d) first few diagrams generated in the modified form of the gap equation (a) with the choice of $\Sigma_{12}$ from (c).
                                     Here, dashed lines represent the interaction potential $v_{0}$ and full lines the matrix elements of the fermionic single-particle propagator $\mathcal{G}_{ij}$
                                     with Nambu notation, while coloured boxes correspond to the pair propagator (\ref{general-pair-propagator-below_Tc}) built up on self-consistent $\mathcal{G}_{ij}$ \cite{footnote-1}.
                                     For compactness, four-wave vectors have not been indicated in the diagrams.}
\label{Figure-1}
\end{center} 
\end{figure}
\noindent
A diagrammatic representation of Eq.~(\ref{modified-gap-equation}) is shown in Fig.~\ref{Figure-1}(a).
Although it contains in principle the same information provided by the original gap equation (\ref{gap-equation}), we shall argue below that the modified form (\ref{modified-gap-equation}) of the gap equation is especially convenient for including in the gap the effects of pairing fluctuations beyond mean field.

We pass now to show that the modified form (\ref{modified-gap-equation}) of the gap equation is equivalent to a \emph{Hugenholtz-Pines condition} for fermion pairs, which generalises to the present context the well-known Hugenholtz-Pines condition for point-like bosons \cite{HP-1959}.
Later in Sections~\ref{sec:G-MB-BCS-BEC} and \ref{sec:numerical-results} we shall see that it is actually due to this equivalence that the effects of pairing fluctuations on the gap parameter can be most readily included in the gap equation.
And this not only in the conventional BCS (weak-coupling) limit, but also throughout the BCS-BEC crossover for which a description in terms of composite bosons applies in the BEC limit.

Alternatively, Eq.~(\ref{modified-gap-equation}) can be considered as a kind of \emph{``tadpole condition''} for the vanishing of insertions with zero momentum, a condition which is known to determine the condensate fraction for the case of point-like bosons \cite{Popov-1983,Popov-1987}.
The analogy between the gap equation for fermions and the tadpole condition for bosons was already pointed out in Ref.~\cite{Pieri-2006}, where this condition was used to deal with composite bosons in the BEC limit.
By the present approach, however, this condition is regarded to apply generically to fermion pairs across the \emph{whole} BCS-BEC crossover and not only in the BEC limit.

\vspace{0.05cm}
\begin{center}
{\bf B. Equivalence of the gap equation with the Hugenholtz-Pines condition for fermion pairs}
\end{center}
\vspace{-0.2cm}

The equivalence of the modified form of the gap equation (\ref{modified-gap-equation}) with the Hugenholtz-Pines condition for fermion pairs is proved as follows, in terms of diagrammatic considerations that hold in the broken-symmetry phase.

\vspace{0.1cm}
\noindent
(i) By construction, the fermionic pairing theory privileges the Fock-like diagram for the anomalous self-energy $\Sigma_{12}$, which is depicted in Fig.~\ref{Figure-1}(b) and given analytically by (minus) the right-hand side of Eq.~(\ref{gap-equation}) \cite{footnote-2}.
This implies that any choice for $\Sigma_{12}$ (at or beyond the mean-field level) must necessarily contain \emph{at least} the Fock-like diagram of Fig.~\ref{Figure-1}(b).
This diagram contributes to Eq.~(\ref{modified-gap-equation}) the factor $-\Delta \, [\mathcal{A}(0) - \mathcal{B}(0)]$, where the (regularised) normal ($\mathcal{A}$) and anomalous ($\mathcal{B}$) particle-particle bubbles in the broken-symmetry phase are defined by \cite{Pistolesi-1996,Andrenacci-2003}:
\begin{eqnarray}
\mathcal{A}(q) & = & - \frac{1}{v_{0}} + \sum_{k} \mathcal{G}_{11}(k+q) \, \mathcal{G}_{22}(k) 
\label{definition-cal_A} \\
\mathcal{B}(q) & = & \sum_{k} \mathcal{G}_{12}(k+q) \, \mathcal{G}_{12}(k) 
\label{definition-cal_B}
\end{eqnarray}
\noindent
with the four-vector notation $q=(\mathbf{q},\Omega_{\nu})$ where $\Omega_{\nu}=2 \pi \nu T$ ($\nu$ integer) is bosonic Matsubara frequency.
This preliminary choice for $\Sigma_{12}$ has also the effect of breaking at the outset the superfluid symmetry of interest \cite{Nozieres-1964}.

\vspace{0.1cm}
\noindent
(ii) Any additional diagrammatic contribution to the anomalous fermionic self-energy $\Sigma_{12}$ (over and above the Fock-like one) is bound to contain at least one anomalous fermionic propagator 
$\mathcal{G}_{12}$ in its skeleton structure.
As an example, the additional contribution to $\Sigma_{12}$ can be taken within the so-called (self-consistent) $t$-matrix approximation shown in Fig.~\ref{Figure-1}(c)
(cf. Appendix~\ref{sec:appendix-A} for a summary of the main features of the $t$-matrix approximation below $T_{c}$).
Each of the anomalous propagators $\mathcal{G}_{12}$ entering this additional contribution to $\Sigma_{12}$ can, in turn, be represented via the identity (\ref{general-identity}) \cite{footnote-3}.

\vspace{0.1cm}
\noindent
(iii) At this point, the total self-energy $\Sigma_{12}$ associated with the identity (\ref{general-identity}) contains both the Fock-like term of Fig.~\ref{Figure-1}(b) (which can be 
set equal to $-\Delta$) and the chosen additional contribution.
In the second case, the above replacement process can go on by a repeated use of the identity (\ref{general-identity}), which at each step gives rise to a term proportional to $\Delta$.
Specifically, Fig.~\ref{Figure-1}(d) shows examples of the diagrammatic terms generated by applying this procedure to $\Sigma_{12}$ of Fig.~\ref{Figure-1}(c), which are
readily recognised as having the topological structure of the Maki-Thompson (MT) and Aslamazov-Larkin (AL) processes.
In general, this is an open-ended process which results in an infinite number of two-particle diagrammatic structures being generated.
In the specific example here considered, one ends up with sequences of MT and AL structures plus a mixed sequence of them, as indeed expected for a conserving approximation \cite{Baym-1962}.

\vspace{0.1cm}
\noindent
(iv) The overall sign of the analytic expression associated with a given two-particle diagrammatic structure (like those shown in Fig.~\ref{Figure-1}(d) or in Fig.~\ref{Figure-2} below) is given by 
$(-1)^{\mathcal{N}_{22} +1}$, where $\mathcal{N}_{22}$ is the number of single-particle propagators $\mathcal{G}_{22}$ that enter the given diagram.

\vspace{0.1cm}
\noindent
(v) When the above considerations are transferred to the right-hand side of the modified form of the gap equation (\ref{modified-gap-equation}), the gap $\Delta$ factors out in all terms in such a way that it can be simplified 
from both sides of the equation.

In this way, Eq.~(\ref{modified-gap-equation}) reduces to the \emph{Hugenholtz-Pines condition} for fermion pairs, in the form:
\begin{equation}
\mathcal{A}(0) - \mathcal{B}(0) + \Sigma^{\mathrm{B}}_{11}(0) - \Sigma^{\mathrm{B}}_{12}(0) = 0 \, .
\label{Hugenholtz-Pines-condition-A-B}
\end{equation}
In this expression, $\Sigma^{\mathrm{B}}_{11}$ and $\Sigma^{\mathrm{B}}_{12}$ correspond to the sequence of diagrams generated as above in the two-particle channel, which act, respectively, as normal and anomalous bosonic-like 
self-energies for the \emph{``bare'' pair propagator} $T_{ij}(q)$ (which is here considered only in the limit $q=0$).
This pair propagator is built in the two-particle channel as a series of ladder diagrams (which are, in turn, derived from the Fock-like diagram of Fig.~\ref{Figure-1}(b) for the anomalous fermionic self-energy 
$\Sigma_{12}$ - cf. Appendix~\ref{sec:appendix-A}) and is given by:
\begin{eqnarray}
\left( 
\begin{array}{cc}
T_{11}(q)  &  T_{12}(q)  \\
T_{21}(q)  &  T_{22}(q)
\end{array} 
\right)
& = & \frac{1}{\mathcal{A}(q) \mathcal{A}(-q) - \mathcal{B}(q)^{2}}
\nonumber \\
& \times &
\left( \!
\begin{array}{cc}
- \mathcal{A}(-q)  &  \mathcal{B}(q)  \\
\mathcal{B}(q)     & - \mathcal{A}(q)
\end{array} 
\! \right) \, .
\label{general-pair-propagator-below_Tc}
\end{eqnarray}
\noindent
Here and in the following, the suffices $(i,j)$ attached to bosonic-like quantities (namely, $T$ and $\Sigma^{\mathrm{B}}$) are identified according to the conventions introduced in Ref.~\cite{Andrenacci-2003},
which relate only indirectly to the Nambu's conventions (\ref{Nambu-notation}) and (\ref{Gorkov-Nambu-equations}) for the single-particle propagators
(cf. Appendix~\ref{sec:appendix-A} for a summary of these conventions).
In terms of the matrix elements $T_{ij}(0)$, the Hugenholtz-Pines condition (\ref{Hugenholtz-Pines-condition-A-B}) for fermion pairs then reads:
\begin{equation}
T^{-1}_{11}(0) - T^{-1}_{12}(0) - \Sigma^{\mathrm{B}}_{11}(0) + \Sigma^{\mathrm{B}}_{12}(0) = 0 \, .
\label{Hugenholtz-Pines-condition-T}
\end{equation}
\noindent
Note that this condition guarantees the \emph{``dressed'' pair propagator} defined by $\bar{T}(q) = [T^{-1}(q) - \Sigma^{\mathrm{B}}(q)]^{-1}$ to remain gapless at $q=0$.
In this way, the value of the thermodynamic gap $\Delta$, which is obtained by solving the modified form of the gap equation in the form of the Hugenholtz-Pines condition (\ref{Hugenholtz-Pines-condition-T}) for fermion pairs, is also related to the dynamical excitations of the systems.
Note also that all our conclusions hold irrespective of the value of the coupling parameter $(k_{F} a_{F})^{-1}$ that spans the BCS-BEC crossover.

This concludes our formal proof that, quite generally, the gap equation that determines the gap parameter is equivalent to a Hugenholtz-Pines condition for fermion pairs within any self-consistent
(or, better, conserving \cite{Baym-1962}) approximation for the single-particle self-energy, with the physical condition that the latter contains at least the Fock-like diagram of Fig.~\ref{Figure-1}(b) for the anomalous self-energy $\Sigma_{12}$, consistently with the pairing theory of superconductivity \cite{BCS-1957,Schrieffer-1964}.

Before concluding this Section, it is worth mentioning an issue that was raised  in Ref.~\cite{Strinati-2015} in a related context, where it was pointed out that a given fermionic conserving approximation results into a gapless approximation for the composite bosons built in terms of the constituent fermions.
In Ref.~\cite{Strinati-2015}, however, no explicit mention was made to the gap equation, so that the two-particle processes resulting from the series of ladder diagrams (that correspond to the bare pair propagator $T$) or from more complex diagrammatic structures (like the series of MT and AL kernels shown above) were considered on equal footing.
The modified form of the gap equation here considered, on the other hand, by its own nature privileges the series of ladder diagrams and thus focuses directly on taking into account more complex diagrammatic structures which act as bosonc-like self-energy corrections just to the series of ladder diagrams.

\section{Implementing Hugenholtz-Pines condition for fermion pairs within the Popov and GMB contributions} 
\label{sec:G-MB-BCS-BEC}
\vspace{-0.3cm}

In the previous Section, we have proven the \emph{formal} equivalence between the modified form of the gap equation and the Hugenholtz-Pines condition for fermion pairs.
The motivation behind this proof was that this equivalence should make it easier to focus directly on (and thus to include) the relevant pairing fluctuations corrections beyond mean field in the gap equation.
Nonetheless, solving numerically the modified form of the gap equation is evidently going to be a quite difficult task, especially if one would keep \emph{all} terms required by a strict implementation of a conserving approximation like in the specific example of Fig.~\ref{Figure-1}.
This implies that, in practice, less demanding (albeit relevant on physical grounds) conditions have to be requested.

In this Section, we implement explicitly the use of the modified form of the gap equation as Hugenholtz-Pines condition for fermion pair, by considering a specific path for the inclusion of pairing fluctuations beyond mean field on the gap parameter.
This path will lead us to consider the long-pending problem about the inclusion of the so-called Gorkov-Melik-Barkhudarov (GMB) correction directly on the gap equation, throughout the BCS-BEC crossover and 
for any temperature in the superfluid phase.
Our approach contrasts (yet duly complements) the original GMB approach of Ref.~\cite{GMB-1961}, where the value of the gap parameter was determined through an instability condition only in the (extreme) BCS limit and at zero temperature.

As it will turn out from the related numerical calculations presented in Section~\ref{sec:numerical-results}, adding the diagrammatic GMB correction (together with an additional Popov diagrammatic correction, see below) on top of the (non-self-consistent) $t$-matrix approximation for the bare pair propagator proves sufficient to account for the gap parameter over the whole BCS-BEC crossover with good accuracy.
In this context, we will be reassured by the agreement obtained when confronting our numerical calculations with the experimental and QMC data that are available in the crossover region of most interest.

\vspace{0.05cm}
\begin{center}
{\bf A. Rephrasing the gap equation at the mean-field level}
\end{center}
\vspace{-0.2cm}

To begin with, it is convenient to rephrase the standard mean-field approach for the gap equation in the form of the modified form of the gap equation (\ref{modified-gap-equation}), as a basis for the inclusion of the relevant pairing-fluctuation corrections over and above the mean field itself.

To this end, as a first step we explicitly verify the identity (\ref{general-identity}) within the mean-field approximation, whereby $\Sigma_{11}(k)=0$ and $\Sigma_{12}(k)=-\Delta^{\mathrm{BCS}}$.
In this case, $\mathcal{G}_{ij}(k) \rightarrow \mathcal{G}_{ij}^{\mathrm{BCS}}(k)$ are given by \cite{FW-1971}:
\begin{eqnarray}
\mathcal{G}_{11}^{\mathrm{BCS}}(k) & = & \frac{u_{\mathbf{k}}^{2}}{i \omega_{n} - E_{\mathbf{k}}} + \frac{v_{\mathbf{k}}^{2}}{i \omega_{n} + E_{\mathbf{k}}}  
\label{BCS-propagator-11} \\
& = &  - \, \mathcal{G}_{22}^{\mathrm{BCS}}(-k) = \mathcal{G}_{0}(k) - \mathcal{G}_{0}(k) \, \Delta^{\mathrm{BCS}} \, \mathcal{G}_{21}^{\mathrm{BCS}}(k)
\nonumber
\end{eqnarray}
\noindent
and
\begin{eqnarray}
\mathcal{G}_{12}^{\mathrm{BCS}}(k) & = & - u_{\mathbf{k}} \, v_{\mathbf{k}} \left( \frac{1}{i \omega_{n} - E_{\mathbf{k}}} - \frac{1}{i \omega_{n} + E_{\mathbf{k}}} \right) 
\label{BCS-propagator-12} \\
& = & \mathcal{G}_{21}^{\mathrm{BCS}}(k) = - \mathcal{G}_{0}(k) \, \Delta^{\mathrm{BCS}} \, \mathcal{G}_{22}^{\mathrm{BCS}}(k) \, .
\nonumber
\end{eqnarray}
\noindent
In the above expressions, $E_{\mathbf{k}} = \sqrt{\xi_{\mathbf{k}}^{2} + (\Delta^{\mathrm{BCS}})^{2}}$ for an isotropic (s-wave) order parameter (or pairing gap) $\Delta^{\mathrm{BCS}}$ within mean field,
\begin{equation}
u_{\mathbf{k}} = \sqrt{\frac{1}{2} \left( 1 + \frac{\xi_{\mathbf{k}}}{E_{\mathbf{k}}} \right)} \,\,\, , \,\,\, v_{\mathbf{k}} = \sqrt{\frac{1}{2} \left( 1 - \frac{\xi_{\mathbf{k}}}{E_{\mathbf{k}}} \right)} \, ,
\label{u-v-BCS}
\end{equation}
\noindent
and $\mathcal{G}_{0}(k) = \left( i \omega_{n} - \xi_{\mathbf{k}} \right)^{-1}$ is the non-interacting fermionic propagator.
By entering the upper lines of the expressions (\ref{BCS-propagator-11}) and (\ref{BCS-propagator-12}) into the identity (\ref{general-identity}), this identity can be verified through simple manipulations.

Within the mean-field approximation, one can further explicitly verify that the second term on the right-hand side of the identity (\ref{general-identity}) acts to cancel a number of undesired terms
that would be present in the first term therein.
This check can be done by using in the identity (\ref{general-identity}) the lower lines of the expressions (\ref{BCS-propagator-11}) and (\ref{BCS-propagator-12}), and expanding the resulting expressions in powers of $\Delta^{\mathrm{BCS}}$.

In addition, within the mean-field approximation the modified form of the gap equation (\ref{modified-gap-equation}) gets considerably simplified since $\Sigma_{12}(k)=-\Delta^{\mathrm{BCS}}$ is a constant. 
With the definitions 
\begin{eqnarray}
A(q) & = & - \frac{1}{v_{0}} + \sum_{k} \mathcal{G}_{11}^{\mathrm{BCS}}(k+q) \, \mathcal{G}_{22}^{\mathrm{BCS}}(k) 
\label{definition-A} \\
B(q) & = & \sum_{k} \mathcal{G}_{12}^{\mathrm{BCS}}(k+q) \, \mathcal{G}_{12}^{\mathrm{BCS}}(k)
\label{definition-B}
\end{eqnarray}
\noindent
which are obtained from Eqs.~(\ref{definition-cal_A}) and (\ref{definition-cal_B}), respectively, with the replacement $\mathcal{G}_{ij} \rightarrow \mathcal{G}_{ij}^{\mathrm{BCS}}$,
the modified form of the gap equation (\ref{modified-gap-equation}) becomes:
\begin{eqnarray}
A(0) - B(0) & = & - \!\! \int \! \frac{d\mathbf{k}}{(2 \pi)^{3}} \left( \frac{1 - 2 f(E_{\mathbf{k}})}{2 E_{\mathbf{k}}} - \frac{m}{\mathbf{k}^{2}} \right) - \frac{m}{4 \pi a_{F}} 
\nonumber \\
& = & 0 
\label{BCS-gap-equation-A_B}
\end{eqnarray} 
\noindent
where $f(E)=(e^{E/T}+1)^{-1}$ is the Fermi function and the regularization condition (\ref{regularization}) has been utilised.
Equation~(\ref{BCS-gap-equation-A_B}) coincides with the standard mean-field equation for $\Delta$ in the case of a contact inter-particle interaction.
This corresponds to the Hugenholtz-Pines condition (\ref{Hugenholtz-Pines-condition-A-B}) for fermion pair with vanishing bosonic-like self-energies $\Sigma^{\mathrm{B}}_{ij}$, consistently with the absence of pairing fluctuations beyond mean field.

The mean-field gap equation (\ref{BCS-gap-equation-A_B}) has to be supplemented by the equation for the density $n$ to determine the chemical potential.
Within the mean-field approximation this equation reads:
\begin{equation}
n = \!\! \int \! \frac{d\mathbf{k}}{(2 \pi)^{3}} \left( 1 - \frac{\xi_{\mathbf{k}}}{E_{\mathbf{k}}} \left(1 - 2 f(E_{\mathbf{k}}) \right) \right) \, .
\label{BCS-density-equation}
\end{equation}

In particular, in the BCS limit $(k_F\, a_F)^{-1} \ll -1$ at zero temperature (whereby $f(E_{\mathbf{k}})=0$), one can use the methods discussed in Refs.~\cite{Popov-1983,Popov-1987} to handle the behaviour of the integrands in Eqs.~(\ref{BCS-gap-equation-A_B}) and (\ref{BCS-density-equation}) in the vicinity of $\xi_{\mathbf{k}} = 0$.
One obtains for the gap equation:
\begin{eqnarray}
- \frac{m}{4 \pi a_{F}} & = & \int \!\! \!\frac{d\mathbf{k}}{(2 \pi)^3} \! \left( \frac{1}{2 E_{\mathbf{k}}} - \frac{m}{\mathbf{k}^{2}} \right)
\nonumber \\
& \simeq & \frac{m \, k_{\mu}}{2 \pi^{2}} \! \left[ \ln \! \left(\frac{8 \mu}{\Delta^{\mathrm{BCS}}_{0}}\right) \! - 2 \right]
\label{gap-equation-BCS-approximation-T=0}
\end{eqnarray}
\noindent
with the wave vector $k_{\mu}$ defined by $\mu = k_{\mu}^{2}/(2m)$ ($\mu > 0$), as well as
\begin{equation}
n \simeq \frac{m \, k_{\mu}}{2 \pi^{2}} \, \mu \left[ \frac{4}{3} + \frac{1}{2} \left( \frac{ \Delta^{\mathrm{BCS}}_{0}}{\mu} \right)^{2} \ln \left(\frac{8 \mu}{\Delta^{\mathrm{BCS}}_{0}}\right)\right] 
\label{density-equation-BCS-approximation-T=0}
\end{equation}
\noindent
for the density.
By solving the above expressions in terms of the coupling parameter $(k_F\, a_F)^{-1}$ (where $a_{F}<0$), they become:
\begin{eqnarray}
\Delta^{\mathrm{BCS}}_{0} & \simeq & \frac{8 \mu}{e^{2}} \, \exp \left( \frac{\pi}{2 k_{\mu} a_{F}} \right) 
\label{final-expression-gap-BCS-approximation-T=0-upper}\\
& \simeq & \frac{8 E_{F}}{e^{2}} \exp \left( \! \frac{\pi}{2 k_{F} a_{F}} \! \right)
\label{final-expression-gap-BCS-approximation-T=0-lower} \\
\frac{\mu}{E_{F}} & \simeq & 1 + \frac{\pi}{8 k_{F} a_{F}} \left( \frac{ \Delta^{\mathrm{BCS}}_{0}}{E_{F}} \right)^{2}
\label{final-expression-density-BCS-approximation-T=0}
\end{eqnarray}
\noindent
where $E_{F}=k_{F}^{2}/(2m)$ is the Fermi energy.
Note that, within mean field, the difference $\mu - E_{F}$ is exponentially small in the coupling parameter $(k_F\, a_F)^{-1}$.
Although this result is sufficient to justify the replacement $\mu \rightarrow E_{F}$ made on the right-hand side of Eq.~(\ref{final-expression-gap-BCS-approximation-T=0-lower}), 
on physical grounds one would have expected the difference $\mu - E_{F}$ to be related to a ``mean-field shift'' and thus to be linear in $k_{F} a_{F}$.
For this to occur, however, pairing-fluctuation corrections need to be included, as shown in subsection~\ref{sec:G-MB-BCS-BEC}-B below.

In the BEC limit whereby $\mu/T \rightarrow - \infty$, on the other hand, it is possible to expand
\begin{eqnarray}
\mathcal{G}_{11}^{\mathrm{BCS}}(k) & = & - \, \mathcal{G}_{22}^{\mathrm{BCS}}(-k)
\nonumber \\ 
& \simeq & \mathcal{G}_{0}(k) - \left( \Delta_{0}^{\mathrm{BCS}} \right)^{2} \, \mathcal{G}_{0}(k)^{2} \, \mathcal{G}_{0}(-k)
\label{approximate-BCS-propagator-11} \\
\mathcal{G}_{12}^{\mathrm{BCS}}(k) & \simeq & \Delta_{0}^{\mathrm{BCS}} \, \mathcal{G}_{0}(k) \, \mathcal{G}_{0}(-k) \, ,
\label{approximate-BCS-propagator-12}
\end{eqnarray}
\noindent
such that the expressions (\ref{definition-A}) and (\ref{definition-B}) with $q=0$ become with the help of the regularization (\ref{regularization}):
\begin{eqnarray}
A(0) & \simeq & - \frac{m}{4 \pi a_{F}} - \sum_{k} \mathcal{G}_{0}(k) \, \mathcal{G}_{0}(-k) + \int \!\! \frac{d\mathbf{k}}{(2 \pi)^3} \, \frac{m}{\mathbf{k}^2}
\nonumber \\
& + & 2 \left( \Delta_{0}^{\mathrm{BCS}} \right)^{2} \sum_{k} \mathcal{G}_{0}(k)^{2} \, \mathcal{G}_{0}(-k)^{2}
\label{approximate-A-BEC_limit} \\
B(0) & \simeq & \left( \Delta_{0}^{\mathrm{BCS}} \right)^{2} \sum_{k} \mathcal{G}_{0}(k)^{2} \, \mathcal{G}_{0}(-k)^{2} \, .
\label{approximate-B-BEC_limit}
\end{eqnarray}
\noindent
Here,
\begin{eqnarray}
& & \!\!\! \sum_{k} \mathcal{G}_{0}(k) \, \mathcal{G}_{0}(-k) - \!\! \int \!\! \frac{d\mathbf{k}}{(2 \pi)^3} \, \frac{m}{\mathbf{k}^2} \simeq - \frac{m}{4 \pi a_{F}} + \! \left( \! \frac{m^{2} a_{F}}{8 \pi} \! \right) \mu_{B}
\nonumber \\
& & \!\! \sum_{k} \mathcal{G}_{0}(k)^{2} \, \mathcal{G}_{0}(-k)^{2} \simeq \left( \! \frac{m^{2} a_{F}}{8 \pi} \! \right)^{2} \left( \frac{4 \pi a_{F}}{m} \right)
\label{special-sums}
\end{eqnarray}
\noindent
where $\mu_{B} = 2 \mu + \epsilon_{0}$ is the chemical potential of the composite bosons that form in this limit (with $\epsilon_{0} = (m a_{F}^{2})^{-1}$ the binding energy of the two-fermion problem).
Entering the approximate results (\ref{special-sums}) into the expressions (\ref{approximate-A-BEC_limit}) and (\ref{approximate-B-BEC_limit}), the modified form of gap equation (\ref{BCS-gap-equation-A_B}) at the mean-field level
becomes eventually:
\begin{equation}
\mu_{B} \simeq \frac{4 \pi (2 a_{F})}{2 m} \, n_{0}
\label{mu_B-vs-n_0-mean-field}
\end{equation}
\noindent
where $n_{0} = \left( \frac{m^{2} a_{F}}{8 \pi} \right) \left( \Delta_{0}^{\mathrm{BCS}} \right)^{2}$ acquires the meaning the \emph{condensate density} \cite{Pieri-2003}.
The result (\ref{mu_B-vs-n_0-mean-field}) further identifies $a_{B} = 2 a_{F}$ as the value of the scattering length $a_{B}$ for the low-energy scattering of composite bosons, at the level of the Born approximation.
We shall see below that the inclusion of pairing fluctuations beyond mean field affects the result (\ref{mu_B-vs-n_0-mean-field}) in two ways, namely, by modifying the value of $a_{B}$ and by adding the contribution of the non-condensate density.

We note, finally, that the approximate expressions (\ref{final-expression-gap-BCS-approximation-T=0-lower}), (\ref{final-expression-density-BCS-approximation-T=0}), and (\ref{mu_B-vs-n_0-mean-field}) could also be recovered from the analytic solution of the mean-field equations (\ref{BCS-gap-equation-A_B}) and (\ref{BCS-density-equation}), as obtained in Ref.~\cite{MPS-1998} at $T=0$.

\vspace{0.05cm}
\begin{center}
{\bf B. $t$-matrix approximation below $T_{c}$}
\end{center}
\vspace{-0.2cm}

The first level of approximation, for the inclusion of pairing fluctuations beyond mean field  in the broken-symmetry phase, is represented by the $t$-matrix approximation, which was studied in Ref.~\cite{Pisani-2004} throughout the BCS-BEC crossover.
This approximation rests on a pair propagator that corresponds to a series of ladder diagrams and is given by the expression (\ref{general-pair-propagator-below_Tc}), whereby $\mathcal{A}$ and $\mathcal{B}$ 
of Eqs.~(\ref{definition-cal_A}) and (\ref{definition-cal_B}) are replaced, respectively, by $A$ and $B$ of Eqs.~(\ref{definition-A}) and (\ref{definition-B}), namely:
\begin{eqnarray}
\left( 
\begin{array}{cc}
T_{11}(q)  &  T_{12}(q)  \\
T_{21}(q)  &  T_{22}(q)
\end{array} 
\right)
& \rightarrow & \frac{1}{A(q) A(-q) - B(q)^{2}}
\nonumber \\
& \times &
\left( \!
\begin{array}{cc}
- A(-q)  &   B(q)  \\
   B(q)  & - A(q)
\end{array} 
\! \right) \, .
\label{pair-propagator-below_Tc}
\end{eqnarray}
\noindent
In addition, in the expressions of $A$ and $B$, $\Delta^{\mathrm{BCS}}$ is replaced by a new value $\Delta$ to be consistently determined.
Within this approximation, the gap equation maintains the formal structure of the mean-field gap equation (\ref{BCS-gap-equation-A_B}), although now the value of the chemical potential therein differs from that obtained by the mean-field density equation (\ref{BCS-density-equation}).
This is because, within the $t$-matrix approximation in the broken-symmetry phase, the density equation reads \cite{Pisani-2004}:
\begin{equation}
n = 2 \sum_{k} e^{i \omega_{n} \eta} \mathcal{G}_{11}(k)
\label{density-equation-t_matrix}
\end{equation}
\noindent
($\eta \rightarrow 0^{+}$ being a positive infinitesimal), where
\begin{eqnarray}
\mathcal{G}_{11}(k) & = & - \mathcal{G}_{22}(-k) 
\label{G_11-generic} \\
& = & \mathcal{G}_{0}(k) + \mathcal{G}_{0}(k) \left[ \Sigma_{11}(k) \mathcal{G}_{11}(k) + \Sigma_{12}(k) \mathcal{G}_{21}(k) \right]
\nonumber 
\end{eqnarray}
\begin{equation}
\mathcal{G}_{12}(k) = \mathcal{G}_{21}(k) = \mathcal{G}_{0}(k) \left[ \Sigma_{11}(k) \mathcal{G}_{12}(k) + \Sigma_{12}(k) \mathcal{G}_{22}(k) \right]
\label{G_12-generic} 
\end{equation}
\noindent
with the following choice of the fermionic self-energy:
\begin{eqnarray}
\Sigma_{11}(k) & = & - \Sigma_{22}(-k) = - \sum_{q} T_{11}(q) \, \mathcal{G}_{22}^{\mathrm{BCS}}(k-q)
\label{Sigma_11-t_matrix} \\
\Sigma_{12}(k) & = & \Sigma_{21}(k) = - \Delta \, .
\label{Sigma_21-t_matrix}
\end{eqnarray}
\noindent
In the above expression, $T_{11}$ is given by Eq.~(\ref{pair-propagator-below_Tc}) and $\mathcal{G}_{22}^{\mathrm{BCS}}$ has still the BCS form (\ref{BCS-propagator-11}) with 
$\Delta^{\mathrm{BCS}}$ replaced by $\Delta$ \cite{footnote-4}.
In addition, $\mathcal{G}_{11}$ of Eq.~(\ref{G_11-generic}) can be conveniently rewritten in the following form \cite{Pisani-2004}:
\begin{equation}
\mathcal{G}_{11}(k) = \left[ \mathcal{G}_{0}^{-1}(k) - \sigma_{11}(k) \right]^{-1}
\label{G_11-convenient_form}
\end{equation}
\noindent
where
\begin{equation}
\sigma_{11}(k) = \Sigma_{11}(k) - \frac{\Delta^{2}}{ \mathcal{G}_{0}^{-1}(-k) - \Sigma_{11}(-k) } \, .
\label{sigma_11}
\end{equation}

Since the gap equation has still the mean-field form (\ref{BCS-gap-equation-A_B}), in the BCS limit $(k_F\, a_F)^{-1} \ll -1$ at zero temperature the result 
$\Delta_{0} \simeq \frac{8 \mu}{e^{2}} \, \exp \left( \frac{\pi}{2 k_{\mu} a_{F}} \right)$ still holds (cf. Eq.~(\ref{final-expression-gap-BCS-approximation-T=0-upper})). 
However, owing to the density equation (\ref{density-equation-t_matrix}), the chemical potential acquires now a linear dependence on the small parameter $k_{F} |a_{F}|$, namely,
\begin{equation}
\frac{\mu}{E_{F}} \simeq 1 + \frac{4}{3 \pi} \, k_{F} a_{F}
\label{chemical-potential-t-matrix-T=0}
\end{equation}
\noindent
instead of the exponential dependence given by Eqs.~(\ref{final-expression-gap-BCS-approximation-T=0-lower}) and (\ref{final-expression-density-BCS-approximation-T=0}).
As a result, the pre-factor on the right-hand side of Eq.~(\ref{final-expression-gap-BCS-approximation-T=0-lower}) gets multiplied by $e^{-1/3}$, such that the expected mean-field result for $\Delta$ 
is not recovered in this limit.
To avoid this shortcoming, in Ref.~\cite{Pisani-2004} a constant fermionic self-energy shift $\Sigma_{0}$ was added to the chemical potential $\mu$ in the mean-field fermionic propagators 
(\ref{BCS-propagator-11}) and (\ref{BCS-propagator-12}) on which the theory is built (and thus also in the gap equation (\ref{BCS-gap-equation-A_B})), where $\Sigma_{0}$ reduces to the mean-field shift 
$2 \pi a_{F} n / m$ in the weak-coupling limit.
In this way, a partial degree of self-consistency is effectively included in the non-self-consistent $t$-matrix approximation.
In the following, we shall instead rely on the Popov approximation introduced in Ref.~\cite{Pieri-2005}, in terms of which a partial degree of self-consistency can be included in a more systematic way
throughout the BCS-BEC crossover.

\vspace{0.05cm}
\begin{center}
{\bf C. Popov contribution to the gap equation}
\end{center}
\vspace{-0.1cm}

The Popov approximation for the BCS-BEC crossover was introduced in Ref.~\cite{Pieri-2005}, to devise a fermionic theory which in the strong-coupling (BEC) limit would reduce to 
the Popov description suitably extended from point-like \cite{Popov-1983,Popov-1987} to composite bosons.
To this end, in Ref.~\cite{Pieri-2005} the form of the gap equation was modified with respect to Eq.~(\ref{BCS-gap-equation-A_B}), in such a way that the associated pair propagator remains gapless at $q=0$.
Here, we recover the gap equation of Ref.~\cite{Pieri-2005} through an alternative route which relies on the modified form (\ref{modified-gap-equation}) of the gap equation, identifying in this way the bosonic-like self-energy $\Sigma^{\mathrm{B}}_{\mathrm{Popov}}$ that corresponds to the Popov approximation.

\begin{figure}[t]
\begin{center}
\includegraphics[width=8.2cm,angle=0]{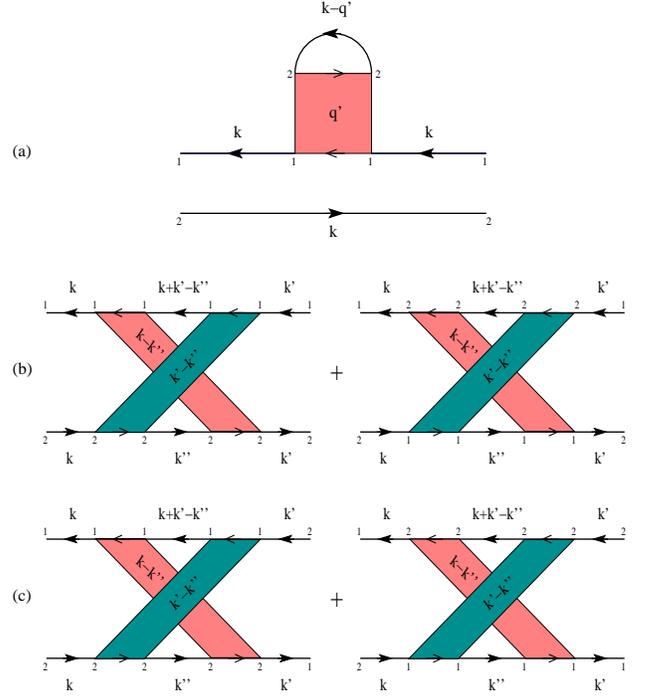}
\caption{(Color online)  Diagrammatic representation of the relevant bosonic-like self-energies in the broken-symmetry phase for $q=0$:
                                      (a) normal component $\Sigma_{\mathrm{Popov}}^{\mathrm{B}}(0)_{11}$ within the Popov approximation (a symmetric dressing of the lower fermionic line
                                            needs also be included);
                                      (b) normal $\Sigma_{\mathrm{GMB}}^{\mathrm{B}}(0)_{11}$ and (c) anomalous $\Sigma_{\mathrm{GMB}}^{\mathrm{B}}(0)_{12}$ components within the GMB  
                                              approximation. 
                                         The coloured boxes correspond to the matrix elements (\ref{pair-propagator-below_Tc}) of the T-matrix.
                                         In all cases, Nambu's indices have explicitly been indicated.}
\label{Figure-2}
\end{center} 
\end{figure}

To this end, in Eq.~(\ref{modified-gap-equation}) we approximate the single-particle fermionic propagators as follows:
\begin{eqnarray}
\mathcal{G}_{11}(k) & \simeq & \mathcal{G}_{11}^{\mathrm{BCS}}(k) + \mathcal{G}_{11}^{\mathrm{BCS}}(k) \, \Sigma_{11}(k) \, \mathcal{G}_{11}^{\mathrm{BCS}}(k) + \cdots
\nonumber \\
\mathcal{G}_{12}(k) & \simeq & \mathcal{G}_{12}^{\mathrm{BCS}}(k) + \cdots \, ,
\label{Popov-approximate-propagators}
\end{eqnarray}
\noindent
and use in addition the forms (\ref{Sigma_11-t_matrix}) and (\ref{Sigma_21-t_matrix}) for the fermionic self-energy.
In this way, Eq.~(\ref{modified-gap-equation}) becomes:
\begin{eqnarray}
- \frac{\Delta}{v_{0}} \simeq - \Delta \, \sum_{k} & & \left[ \mathcal{G}_{11}^{\mathrm{BCS}}(k) \, \mathcal{G}_{22}^{\mathrm{BCS}}(k) 
                                                                    - \mathcal{G}_{12}^{\mathrm{BCS}}(k) \, \mathcal{G}_{12}^{\mathrm{BCS}}(k) \right.
\nonumber \\
                                                                 & & \left. + \, 2 \,  \mathcal{G}_{11}^{\mathrm{BCS}}(k)^{2} \, \mathcal{G}_{22}^{\mathrm{BCS}}(k) \, \Sigma_{11}(k) \right]
\label{Popov-gap-equation-I}
\end{eqnarray}
\noindent
at the lowest significant order in $\Sigma_{11}$.
With the definitions (\ref{definition-A}) and (\ref{definition-B}), Eq.~(\ref{Popov-gap-equation-I}) can then be cast in the form:
\begin{equation}
A(0) - B(0) + \Sigma_{\mathrm{Popov}}^{\mathrm{B}}(0)_{11} = 0
\label{Popov-gap-equation-II}
\end{equation}
\noindent
where $\Sigma_{\mathrm{Popov}}^{\mathrm{B}}(0)_{11}$ is the $q=0$ value of the Popov bosonic-like self-energy in broken symmetry, given by the expression \cite{Pieri-2005}
\begin{small}
\begin{eqnarray}
& & \hspace{-0.5cm} \Sigma_{\mathrm{Popov}}^{\mathrm{B}}(0)_{11} = 2 \sum_{k} \mathcal{G}_{11}^{\mathrm{BCS}}(k)^{2} \, \mathcal{G}_{22}^{\mathrm{BCS}}(k) \, \Sigma_{11}(k)
\nonumber \\
& = & - 2 \sum_{k,q'}  \mathcal{G}_{11}^{\mathrm{BCS}}(k)^{2} \, \mathcal{G}_{22}^{\mathrm{BCS}}(k) \, \mathcal{G}_{22}^{\mathrm{BCS}}(k-q') \, T_{11}(q')
\label{Popov-bosonic-self-energy}
\end{eqnarray}
\end{small}
\noindent
\hspace{-0.27cm} and depicted diagrammatically in Fig.~\ref{Figure-2}(a) \cite{footnote-4}.

Note that Eq.~(\ref{Popov-gap-equation-II}) has the form of the Hugenholtz-Pines condition (\ref{Hugenholtz-Pines-condition-A-B}) for fermion pairs, which now contains a ``normal'' (diagonal) bosonic-like self-energy in contrast with the mean-field and the $t$-matrix counterparts.
[In the following, we shall sometimes refer to the Popov bosonic-like self-energy $\Sigma_{\mathrm{Popov}}^{\mathrm{B}}(0)_{11}$ that enters the gap equation 
(\ref{Popov-gap-equation-II}) simply as $\Sigma_{\mathrm{Popov}}^{\mathrm{B}}$ \cite{footnote-5}.]
The Hugenholtz-Pines condition (\ref{Popov-gap-equation-II}), which includes the Popov contribution, has to be solved in conjunction with the density equation, which in the present approximation needs to be appropriately modified with respect to Eq.~(\ref{density-equation-t_matrix}), as it will discussed in subsection~\ref{sec:numerical-results}-A below.

The expression (\ref{Popov-bosonic-self-energy}) for the Popov bosonic-like self-energy can be calculated analytically both in the BCS (weak-coupling) and BEC (strong-coupling) limits.
In the BCS limit, one finds $\Sigma_{\mathrm{Popov}}^{\mathrm{B}} =- m k_{\mu}/(6 \pi^{2})$, which cancels the spurious factor $e^{-1/3}$ that affects the expression
of $\Delta_{0}^{\mathrm{BCS}}$ within the $t$-matrix approximation of subsection~\ref{sec:G-MB-BCS-BEC}-B.
Dealing with the expression (\ref{Popov-bosonic-self-energy}) in the BEC limit is somewhat more involved and will be discussed in detail in Appendix~\ref{sec:appendix-B}.

We remark that, in the limit $T \rightarrow T_{c}^{-}$ whereby $\Delta \rightarrow 0$, $A(q)$ reduces to the inverse of the pair propagator $\Gamma_{0}(q)$ in the normal phase, $B(q) \rightarrow 0$, 
and $\mathcal{G}_{11}^{\mathrm{BCS}} \rightarrow \mathcal{G}_{0}$.
In this limit, the Hugenholtz-Pines condition (\ref{Popov-gap-equation-II}) then reduces to the equation that determines the critical temperature $T_{c}$ within the Popov approximation \cite{Pisani-2018}.

\vspace{0.05cm}
\begin{center}
{\bf D. GMB contribution to the gap equation}
\end{center}
\vspace{-0.2cm}

In the original GMB paper \cite{GMB-1961}, the gap parameter $\Delta_{0}$ at zero temperature was calculated in weak coupling only (whereby $\mu=E_{F}$), by searching for the singularities in the complex energy plane of the pair propagator from the normal phase.
Accordingly, a pole was found to occur in the upper-half plane at an energy equal to the BCS gap given by the right-hand side of Eq.~(\ref{final-expression-gap-BCS-approximation-T=0-lower}) divided by
$(4e)^{1/3} \simeq 2.2$.
Since in Ref.~\cite{GMB-1961} the same factor was found also to reduce the value of the critical temperature $T_{c}$ with respect to the BCS value, the ratio $\Delta_{0}/T_{c} = \pi / e^{\gamma}$ was not modified with respect to the BCS value (where $\gamma$ is Euler's constant).

Similarly to what was done for the calculation of $T_{c}$, in the literature the GMB correction for $\Delta_{0}$ in weak coupling was often attributed to screening effects, owing to the occurrence of a particle-hole bubble in its expression.
As mentioned in the Introduction, effects of ``medium polarization'' at zero temperature have been studied also for superfluid nuclear and neutron matter, in terms of a gap equation with a suitably screened inter-particle interaction \cite{Cao-2006}.
However, no consideration was given in that context to the BCS-BEC crossover.
Extension of the GMB correction to the BCS-BEC crossover was instead considered in Ref.~\cite{Chen-2016}, where the characteristic approximations of the (extreme) BCS limit, that were exploited
in the original GMB paper \cite{GMB-1961}, have however not been released even upon approaching the BEC limit.

In the following, we shall rely on the modified form (\ref{modified-gap-equation}) of the gap equation in order to include the GMB contribution, which will naturally lead us to recover the Hugenholtz-Pines condition 
(\ref{Hugenholtz-Pines-condition-A-B}) for fermion pairs with suitably identified bosonic-like self-energies $\Sigma^{\mathrm{B}}_{\mathrm{GMB}}(0)_{11}$ and 
$\Sigma^{\mathrm{B}}_{\mathrm{GMB}}(0)_{12}$.
In this way, we will be able to extend the GMB contribution to the whole BCS-BEC crossover, thereby relaxing the approximations characteristic of the (extreme) BCS limit which completely loose their meaning 
when spanning the BCS-BEC crossover.
In the following, however, no attempt will be made to cast the underlying fermionic theory at the level of a fully conserving approximation, and not even to make it self-consistent at the present level.
These additional features, in fact, would be extremely hard to handle, either by implementing them numerically or by deriving from them reliable analytic results in the BCS and BEC limits.

To this end, we adopt the following approximate choice for the off-diagonal fermionic self-energies to be entered in Eq.~(\ref{modified-gap-equation}):
\begin{eqnarray}
\Sigma_{12}(k) = \Sigma_{21}(k) & \simeq & - \Delta + \sum_{k' k''} T_{11}(k-k'') \, T_{11}(k'-k'') 
\nonumber \\
& \times & \mathcal{G}_{11}(k+k'-k'') \, \mathcal{G}_{12}(k') \, \mathcal{G}_{22}(k'')
\label{fermionic-self-energy-GMB} 
\end{eqnarray}
\noindent
where we further approximate
\begin{eqnarray}
\mathcal{G}_{12}(k') = \mathcal{G}_{21}(k') & \simeq & \mathcal{G}_{11}(k') \, (-\Delta) \, \mathcal{G}_{22}(k') 
\nonumber \\
& - & \mathcal{G}_{12}(k') \, (-\Delta) \, \mathcal{G}_{21}(k')
\label{approximate-G_12}
\end{eqnarray}
\noindent
with the use of the identity (\ref{general-identity}). 
In the expressions (\ref{fermionic-self-energy-GMB}) and (\ref{approximate-G_12}), all fermionic propagators $\mathcal{G}_{ij}$ are taken of the mean-field form 
$\mathcal{G}_{ij}^{\mathrm{BCS}}$ given by Eqs.~(\ref{BCS-propagator-11}) and (\ref{BCS-propagator-12}), with $\Delta^{\mathrm{BCS}}$ replaced by a new value $\Delta$ to be consistently determined \cite{footnote-4}.
In addition, the elements $T_{ij}(q)$ of the many-particle T-matrix are meant to have the approximate form (\ref{pair-propagator-below_Tc}) with the bubbles $A$ and $B$ given by the expressions 
(\ref{definition-A}) and (\ref{definition-B}), respectively. 

Entering the approximate expressions (\ref{fermionic-self-energy-GMB}) and (\ref{approximate-G_12}) into the modified form (\ref{modified-gap-equation}) of the gap equation yields eventually the expression:
\begin{eqnarray}
& & \frac{1}{v_{0}} = \sum_{k} \left[ \mathcal{G}_{11}^{\mathrm{BCS}}(k) \mathcal{G}_{22}^{\mathrm{BCS}}(k) - \mathcal{G}_{12}^{\mathrm{BCS}}(k) \mathcal{G}_{12}^{\mathrm{BCS}}(k) \right]
\nonumber \\
& + & \sum_{k k' k''} T_{11}(k-k'') \, T_{11}(k'-k'') \, \mathcal{G}_{11}^{\mathrm{BCS}}(k+k'-k'') \, \mathcal{G}_{22}^{\mathrm{BCS}}(k'')
\nonumber \\
& \times & \left[ \mathcal{G}_{11}^{\mathrm{BCS}}(k) \mathcal{G}_{22}^{\mathrm{BCS}}(k) - \mathcal{G}_{12}^{\mathrm{BCS}}(k) \mathcal{G}_{12}^{\mathrm{BCS}}(k) \right]
\nonumber \\
& \times & \left[ \mathcal{G}_{11}^{\mathrm{BCS}}(k') \mathcal{G}_{22}^{\mathrm{BCS}}(k') - \mathcal{G}_{12}^{\mathrm{BCS}}(k') \mathcal{G}_{12}^{\mathrm{BCS}}(k') \right] \, .
\label{GMB-gap-equation-I}
\end{eqnarray}
\noindent

We are thus led to introduce the quantities
\begin{widetext}
\begin{eqnarray}
\Sigma_{\mathrm{GMB}}^{\mathrm{B}}(0)_{11} & = &
\sum_{k k' k''} T_{11}(k-k'') \, T_{11}(k'-k'') \, \mathcal{G}_{11}^{\mathrm{BCS}}(k+k'-k'') \, \mathcal{G}_{22}^{\mathrm{BCS}}(k'')
\nonumber \\
& \times & \left[ \mathcal{G}_{11}^{\mathrm{BCS}}(k) \, \mathcal{G}_{22}^{\mathrm{BCS}}(k) \, \mathcal{G}_{11}^{\mathrm{BCS}}(k') \, \mathcal{G}_{22}^{\mathrm{BCS}}(k') 
+ \mathcal{G}_{12}^{\mathrm{BCS}}(k) \, \mathcal{G}_{12}^{\mathrm{BCS}}(k) \, \mathcal{G}_{12}^{\mathrm{BCS}}(k') \, \mathcal{G}_{12}^{\mathrm{BCS}}(k') \right]
\label{GMB-bosonic-self-energy_11} 
\end{eqnarray}

\begin{eqnarray}
\Sigma_{\mathrm{GMB}}^{\mathrm{B}}(0)_{12} & = & 
\sum_{k k' k''} T_{11}(k-k'') \, T_{11}(k'-k'') \, \mathcal{G}_{11}^{\mathrm{BCS}}(k+k'-k'') \, \mathcal{G}_{22}^{\mathrm{BCS}}(k'')
\nonumber \\
& \times & \left[ \mathcal{G}_{11}^{\mathrm{BCS}}(k) \, \mathcal{G}_{22}^{\mathrm{BCS}}(k) \, \mathcal{G}_{12}^{\mathrm{BCS}}(k') \, \mathcal{G}_{12}^{\mathrm{BCS}}(k') 
+ \mathcal{G}_{12}^{\mathrm{BCS}}(k) \, \mathcal{G}_{12}^{\mathrm{BCS}}(k) \, \mathcal{G}_{11}^{\mathrm{BCS}}(k') \, \mathcal{G}_{22}^{\mathrm{BCS}}(k') \right] 
\label{GMB-bosonic-self-energy_12} \, ,
\end{eqnarray}
\end{widetext}
\noindent
which represent the $q=0$ values of the``normal'' (diagonal)  and ``anomalous'' (off-diagonal) bosonic-like self-energy within the GMB approximation in the broken-symmetry phase, 
as depicted diagrammatically in Figs.~\ref{Figure-2}(b) and \ref{Figure-2}(c), respectively.
With these definitions and recalling Eqs.~(\ref{definition-A}) and (\ref{definition-B}), the condition (\ref{GMB-gap-equation-I}) for the gap $\Delta$ acquires the form of the Hugenholtz-Pines condition 
(\ref{Hugenholtz-Pines-condition-A-B}) for fermion pairs, namely,
\begin{equation}
A(0) - B(0) + \Sigma_{\mathrm{GMB}}^{\mathrm{B}}(0)_{11} - \Sigma_{\mathrm{GMB}}^{\mathrm{B}}(0)_{12} = 0 \, ,
\label{GMB-gap-equation-II}
\end{equation}
\noindent
which this time contains the anomalous (off-diagonal) bosonic-like self-energy besides the ÒnormalÓ (diagonal) one when compared to the corresponding Popov result (\ref{Popov-gap-equation-II}).
[In the following, we shall sometimes refer to the difference $\Sigma_{\mathrm{GMB}}^{\mathrm{B}}(0)_{12} - \Sigma_{\mathrm{GMB}}^{\mathrm{B}}(0)_{11}$ that enters the gap equation (\ref{GMB-gap-equation-II}) 
simply as $\Sigma_{\mathrm{GMB}}^{\mathrm{B}}$ \cite{footnote-5}.]

The Hugenholtz-Pines condition (\ref{GMB-gap-equation-II}) holds for all temperatures in the broken-symmetry phase and for all couplings throughout the BCS-BEC crossover.
The numerical solution of Eq.~(\ref{GMB-gap-equation-II}), in conjunction with that of the density equation, will be considered in Section~\ref{sec:numerical-results}.
Here, we focus instead on the analytic results that can be obtained in the (extreme) BCS and BEC limits.

We first consider the (extreme) BCS limit at zero temperature, where the original GMB result of Ref.~\cite{GMB-1961} for the gap parameter $\Delta_{0}$ ought to be recovered from Eq.~(\ref{GMB-gap-equation-II}).
To this end, it is convenient to consider directly the form (\ref{GMB-gap-equation-I}) of the gap equation and adopt therein the following simplifying assumptions that hold in this limit:

\vspace{0.1cm}
\noindent
(i) Approximate $T_{11}(q) \rightarrow 4 \pi a_{F} / m$, similarly to what is done in the normal phase above $T_{c}$ \cite{Pisani-2018} (cf. also Fig.~\ref{Figure-3} below).

\vspace{0.1cm}
\noindent
(ii) As a consequence, the three sums over the four-vectors $(k,k',k'')$ in the second term on the right-hand side of Eq.~(\ref{GMB-gap-equation-I}) get completely decoupled from each other.
Care should, however, be exerted in restoring the convergence of the overall expression, which would be lost by the mere replacement made before in (i).
To this end, we can make a compensating replacement and regularise both sums over $k$ and $k'$ in the following way:
\begin{eqnarray}
& - & \sum_{k} \left[ \mathcal{G}_{11}^{\mathrm{BCS}}(k) \mathcal{G}_{22}^{\mathrm{BCS}}(k) - \mathcal{G}_{12}^{\mathrm{BCS}}(k) \mathcal{G}_{12}^{\mathrm{BCS}}(k) \right]
\label{compensating-replacement} \\
& = & \!\! \int \! \frac{d\mathbf{k}}{(2 \pi)^{3}} \! \left( \! \frac{1 - 2 f(E_{\mathbf{k}})}{2 E_{\mathbf{k}}} \! \right)  \rightarrow 
\int \! \frac{d\mathbf{k}}{(2 \pi)^{3}} \! \left( \! \frac{1 - 2 f(E_{\mathbf{k}})}{2 E_{\mathbf{k}}} - \frac{m}{\mathbf{k}^{2}} \! \right) 
\nonumber
\end{eqnarray}
\noindent
where $f(E_{\mathbf{k}}) \rightarrow 0$ in the zero-temperature limit.

\vspace{0.1cm}
\noindent
(iii) The sum over $k''$ is instead handled as follows:
\begin{eqnarray}
& & \sum_{k''} \mathcal{G}_{11}^{\mathrm{BCS}}(k+k'-k'') \, \mathcal{G}_{22}^{\mathrm{BCS}}(k'') 
\nonumber \\
& = & - \sum_{k''} \mathcal{G}_{11}^{\mathrm{BCS}}(k'') \, \mathcal{G}_{11}^{\mathrm{BCS}}(k''-k-k') 
\nonumber \\
& \simeq & - \sum_{k''} \mathcal{G}_{0}(k'')  \, \mathcal{G}_{0}(k''-k-k') \equiv \chi_{\mathrm{ph}}(k+k')
\label{particle-hole-bubble}
\end{eqnarray}
\noindent
where $\mathcal{G}_{0}$ is the non-interacting propagator and $\chi_{\mathrm{ph}}$ the corresponding particle-hole bubble of the normal phase.
In addition, in this bubble set $\omega_{n}+\omega_{n'}=0$, take the wave vectors $\mathbf{k}$ and $\mathbf{k'}$ on a Fermi sphere of radius $k_{F}$, and perform an averaging over 
their relative angle.
This is because the terms within square brackets in Eq.~(\ref{GMB-gap-equation-I}) are strongly peaked at $|\mathbf{k}|=|\mathbf{k'}|=k_{\mu}$ and $\omega_{n}=\omega_{n'}=0$.
The result is \cite{GMB-1961}:
\begin{equation}
\chi_{\mathrm{ph}}(k+k') \rightarrow \bar{\chi}_{\mathrm{ph}}(0) = - N_{0} \ln \left(4e\right)^{1/3} 
\label{approximate-ph-bubble}
\end{equation}
\noindent
where $N_{0} = m k_{F} / (2 \pi^{2})$ is the single-particle density of states (per spin component) at the Fermi level.

\vspace{0.1cm}
\noindent
(iv) Grouping together the above results (i)-(iii) in the form (\ref{GMB-gap-equation-I}) of the gap equation yields approximately:
\begin{eqnarray}
& & \frac{m}{4 \pi a_{F}} + \int \! \frac{d\mathbf{k}}{(2 \pi)^{3}} \! \left( \! \frac{1 - 2 f(E_{\mathbf{k}})}{2 E_{\mathbf{k}}} - \frac{m}{\mathbf{k}^{2}} \! \right) 
\label{GMB-gap-equation-BCS-limit-I} \\
& + & \left[ \frac{4 \pi a_{F}}{m} \int \! \frac{d\mathbf{k}}{(2 \pi)^{3}} \! \left( \! \frac{1 - 2 f(E_{\mathbf{k}})}{2 E_{\mathbf{k}}} - \frac{m}{\mathbf{k}^{2}} \! \right) \right]^{2} \bar{\chi}_{\mathrm{ph}}(0) = 0 \, .
\nonumber
\end{eqnarray}

\vspace{0.1cm}
\noindent
(v) The expression (\ref{GMB-gap-equation-BCS-limit-I}) can be further simplified by noting that, according to the mean-field result (\ref{BCS-gap-equation-A_B}),
\begin{equation}
\frac{4 \pi a_{F}}{m} \int \! \frac{d\mathbf{k}}{(2 \pi)^{3}} \! \left( \! \frac{1 - 2 f(E_{\mathbf{k}})}{2 E_{\mathbf{k}}} - \frac{m}{\mathbf{k}^{2}} \! \right) = -1 \, .
\label{BCS-result}
\end{equation}
\noindent
In this a way, Eq.~(\ref{GMB-gap-equation-BCS-limit-I}) becomes eventually:
\begin{equation}
\frac{m}{4 \pi a_{F}} + \int \! \frac{d\mathbf{k}}{(2 \pi)^{3}} \! \left( \! \frac{1 - 2 f(E_{\mathbf{k}})}{2 E_{\mathbf{k}}} - \frac{m}{\mathbf{k}^{2}} \! \right) + \, \bar{\chi}_{\mathrm{ph}}(0) = 0 \, .
\label{GMB-gap-equation-BCS-limit-II} 
\end{equation}

In particular, in the zero-temperature limit (whereby $f(E_{\mathbf{k}}) \rightarrow 0$) use of the result (\ref{gap-equation-BCS-approximation-T=0}) with $\mu = E_{F}$ and of Eq.~(\ref{approximate-ph-bubble}) 
for $\bar{\chi}_{\mathrm{ph}}(0)$ brings Eq.~(\ref{GMB-gap-equation-BCS-limit-II}) to the form:
\begin{equation}
\frac{m}{4 \pi a_{F}} + N_{0} \left[ \ln \left( \frac{8 E_{F}}{\Delta_{0}^{\mathrm{GMB}}} \right) - 2 - \ln \left(4e\right)^{1/3} \right] = 0 \, .
\label{GMB-gap-equation-BCS-limit-III}
\end{equation}
\noindent
From this expression the result of Ref.~\cite{GMB-1961} readily follows, namely,
\begin{equation}
\Delta^{\mathrm{GMB}}_{0} = \frac{8 E_{F}}{e^{2} \left(4e\right)^{1/3}} \exp \left\{ \! \frac{\pi}{2 k_{F} a_{F}} \! \right\} = \frac{\Delta^{\mathrm{BCS}}_{0}}{\left(4e\right)^{1/3}} \, .
\label{GMB-Delta-0}
\end{equation}
\noindent

From the way it was derived, it is clear that Eq.~(\ref{GMB-gap-equation-BCS-limit-II}) holds under the specific approximations that are valid only in the (extreme) BCS limit when $(k_{F} a_{F})^{-1} \ll -1$.
Accordingly, one is not justified to consider Eq.~(\ref{GMB-gap-equation-BCS-limit-II}) valid over to the whole BCS-BEC crossover in the broken-symmetry phase for arbitrary values of $(k_{F} a_{F})^{-1}$, as it was done in Ref.~\cite{Chen-2016}.
This is because the very first approximation (i) above, about taking $T_{11}(q) \simeq$ constant independent of wave vector and frequency, is bound to fail away from the (extreme) BCS limit.
This crucial point was recently emphasized for the normal phase in Ref.~\cite{Pisani-2018}, where a proper way to handle the GMB contribution for determining the critical temperature $T_{c}$ throughout 
the BCS-BEC crossover was discussed in detail.

\begin{figure}[t]
\begin{center}
\includegraphics[width=8.8cm,angle=0]{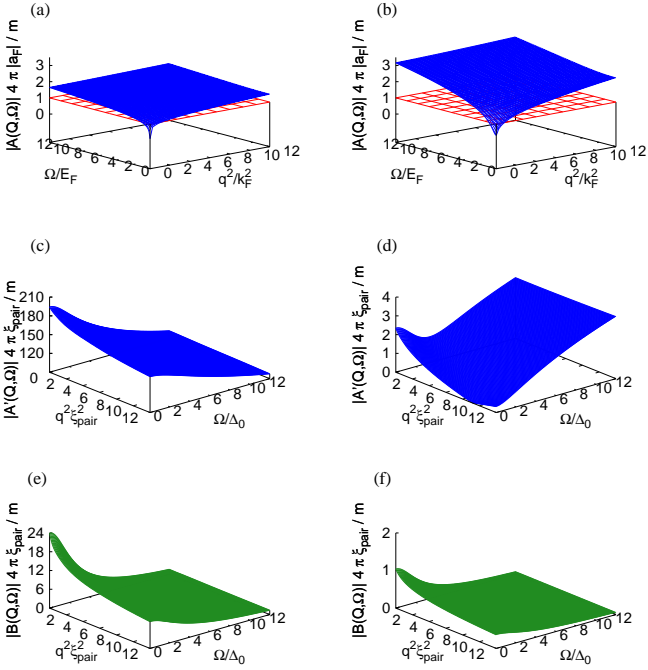}
\caption{(Color online) The magnitudes of $A(\mathbf{q},\Omega_{\nu})$ (upper panels), $A'(\mathbf{q},\Omega_{\nu})$ (middle panels), and $B(\mathbf{q},\Omega_{\nu})$ 
                                    (lower panels) at zero temperature are shown vs $\mathbf{q}^{2}$ and $\Omega_{\nu}$ for the coupling values $(k_{F} a_{F})^{-1} = -3.0$ (left panels) and 
                                    $(k_{F} a_{F})^{-1} = -1.0$ (right panels).
                                    In each case, appropriate normalisations of length and energy are utilised.
                                    In these plots, the mean-field values of $\Delta_{0}$ and $\mu_{0}$ at $T=0$ have been used.}
\label{Figure-3}
\end{center} 
\end{figure}

We can explicitly verify numerically to what extent the approximation $T_{11}(q) \simeq$ constant holds along the BCS-BEC crossover in the broken-symmetry phase.
This is done by plotting the magnitudes of $A(\mathbf{q},\Omega_{\nu})$ and $B(\mathbf{q},\Omega_{\nu})$ at zero temperature for couplings in the extreme BCS limit and at the boundary 
between the BCS and the crossover regimes. 
This is shown in Fig.~\ref{Figure-3}, where the magnitude of $A(\mathbf{q},\Omega_{\nu})$ is plotted in units of $m/(4 \pi |a_{F}|)$ while the magnitude of $B(\mathbf{q},\Omega_{\nu})$ (as well as of $A'(\mathbf{q},\Omega_{\nu})$, where $A'(q) = A(q) + m/(4 \pi a_{F})$ - cf. Eq.~(\ref{definition-A-regularized}) below) is plotted in units of $m/(4 \pi \xi_{\mathrm{pair}})$. 
Here, $\xi_{\mathrm{pair}}$ is the Cooper pair size at zero temperature \cite{Pistolesi-1994}, for which we have used the values $k_{F} \xi_{\mathrm{pair}} = (72.74,3.39)$ for the couplings 
$(k_{F} a_{F})^{-1} = (-3.0,-1.0)$, in the order \cite{MPS-1998}.
Note that for $B$ and $A'$ the gap $\Delta_{0}$ at zero temperature (instead of the Fermi energy $E_{F}$ like for $A$) is used as the unit of energy.
From these plots we conclude that only in the extreme BCS limit $(k_{F} a_{F})^{-1} \ll -1$ can $|A(\mathbf{q},\Omega_{\nu})|$ be considered constant (and equal to $m/(4 \pi |a_{F}|)$) 
over a large portion of the $\mathbf{q}$-$\Omega_{\nu}$ plane (while $|B(\mathbf{q},\Omega_{\nu})|$ is essentially negligible in this regime).

That the particle-hole bubble (\ref{particle-hole-bubble}) is not bound to enter the GMB version (\ref{GMB-gap-equation-II}) of the Hugenholtz-Pines condition away from the BCS regime can be also confirmed by considering the opposite BEC regime, where an analytic calculation of the GMB bosonic-like self-energies $\Sigma_{\mathrm{GMB}}^{\mathrm{B}}(0)_{11}$ and $\Sigma_{\mathrm{GMB}}^{\mathrm{B}}(0)_{12}$ is also possible.
To this end, it is convenient to consider directly the expression for the difference $\Sigma_{\mathrm{GMB}}^{\mathrm{B}}(0)_{11} - \Sigma_{\mathrm{GMB}}^{\mathrm{B}}(0)_{12}$ given by the last term on the right-hand side of Eq.~(\ref{GMB-gap-equation-I}), in which one can make use of the formal identity
\begin{eqnarray}
& & \mathcal{G}_{11}^{\mathrm{BCS}}(k) \mathcal{G}_{22}^{\mathrm{BCS}}(k) - \mathcal{G}_{12}^{\mathrm{BCS}}(k) \mathcal{G}_{12}^{\mathrm{BCS}}(k)
\nonumber \\
& = & \frac{1}{i \omega_{n} - E_{\mathbf{k}}} \, \frac{1}{i \omega_{n} + E_{\mathbf{k}}} = - \, \tilde{\mathcal{G}}_{0}(k) \, \tilde{\mathcal{G}}_{0}(-k) \, ,
\label{identity-BCS-non_interacting}
\end{eqnarray}
\noindent
where $\Delta^{\mathrm{BCS}} \rightarrow \Delta$ and
\noindent
\begin{equation}
\tilde{\mathcal{G}}_{0}(k) = \left( i \omega_{n} - E_{\mathbf{k}} \right)^{-1}
\label{modified-non-interacting-G}
\end{equation}
\noindent
has the form of the non-interacting fermionic propagator $\mathcal{G}_{0}(k)$ with $\xi_{\mathbf{k}}$ replaced by $E_{\mathbf{k}} = \sqrt{\xi_{\mathbf{k}}^{2} + \Delta^{2}}$ .
In this way, we can rewrite the last term on the right-hand side of Eq.~(\ref{GMB-gap-equation-I}) in the compact form:
\begin{eqnarray}
& & \Sigma_{\mathrm{GMB}}^{\mathrm{B}}(0)_{11} - \Sigma_{\mathrm{GMB}}^{\mathrm{B}}(0)_{12}
\nonumber \\
& = & \sum_{k k' k''} T_{11}(k-k'') \, T_{11}(k'-k'') \, \mathcal{G}_{11}^{\mathrm{BCS}}(k+k'-k'') \, 
\nonumber \\
& \times & \mathcal{G}_{22}^{\mathrm{BCS}}(k'') \, \tilde{\mathcal{G}}_{0}(k) \, \tilde{\mathcal{G}}_{0}(-k) \, \tilde{\mathcal{G}}_{0}(k') \, \tilde{\mathcal{G}}_{0}(-k')
\nonumber \\
& = &  \sum_{k p q} T_{11}(p) \, T_{11}(q) \, \mathcal{G}_{11}^{\mathrm{BCS}}(k+q) \, \mathcal{G}_{22}^{\mathrm{BCS}}(k-p)
\nonumber \\
& \times & \tilde{\mathcal{G}}_{0}(k) \, \tilde{\mathcal{G}}_{0}(-k) \, \tilde{\mathcal{G}}_{0}(k+q-p) \, \tilde{\mathcal{G}}_{0}(-k-q+p)
\label{computationally-convenient-Sigma_GMB} 
\end{eqnarray}
\noindent
where we have introduced the bosonic variables $p = k - k''$ and $q = k' - k''$.
In the BEC (strong-coupling) limit we are interested in, whereby $\mu/T \rightarrow - \infty$, the binding energy $\epsilon_{0}$ of the two-fermion problem is much larger than the gap 
$\Delta$ and the temperatures of interest, which are of the order of $T_{c}$.
Under these circumstances, the expression (\ref{computationally-convenient-Sigma_GMB}) will be evaluated analytically in Appendix~\ref{sec:appendix-B}, where it will be explicitly verified
that no remnant of the particle-hole bubble (\ref{particle-hole-bubble}) survives in the BEC limit.

\section{Numerical strategies and results} 
\label{sec:numerical-results}
\vspace{-0.2cm}

In this Section, we obtain numerically the solution of the Hugenholtz-Pines condition for fermion pairs, in the form of Eqs.~(\ref{Popov-gap-equation-II}) or (\ref{GMB-gap-equation-II}) to include separately the
Popov or GMB contribution, or else in the form
\begin{equation}
A(0) - B(0) + \Sigma_{\mathrm{Popov}}^{\mathrm{B}}(0)_{11} + \Sigma_{\mathrm{GMB}}^{\mathrm{B}}(0)_{11} - \Sigma_{\mathrm{GMB}}^{\mathrm{B}}(0)_{12} = 0
\label{Popov_plus_GMB-HP}
\end{equation}
\noindent
to include both contributions simultaneously.
These solutions will be determined as a function of temperature \emph{and\/} coupling throughout the BCS-BEC crossover, in conjunction with the solution of the density equation (\ref{density-equation-t_matrix}).
As a test on the accuracy of our numerical calculations, we will also recover numerically the limiting behaviours of the bosonic-like self-energies $\Sigma_{\mathrm{Popov}}^{\mathrm{B}}$ and 
$\Sigma_{\mathrm{GMB}}^{\mathrm{B}}$ that can be obtained analytically in the BCS and BEC regimes.
This is especially important for the GMB contribution in the BCS limit, for which we will recover numerically the expected result obtained through a different procedure in the original GMB paper \cite{GMB-1961}.
In addition, our numerical results will be compared with available experimental data obtained with ultra-cold Fermi gases and with QMC calculations, as well as with alternative diagrammatic calculations.

\vspace{0.05cm}
\begin{center}
{\bf A. Numerical strategies below $T_{c}$}
\end{center}
\vspace{-0.2cm}

The numerical procedure, to solve the Hugenholtz-Pines condition for fermion pairs with the Popov and/or the GMB contributions, takes advantage of the experience developed in Ref.~\cite{Pisani-2018}, where the critical temperature $T_{c}$ was approached from the normal phase throughout the BCS- BEC crossover.
In that reference, it was found necessary to introduce a (partial) degree of \emph{self-consistency in the pair propagator\/}, in order to avoid entering a temperature regime (below the critical temperature obtained in the absence of the Popov and/or GMB corrections) where the pair propagator itself would diverge at $q=0$.
Here, we adopt a similar strategy also in the broken-symmetry phase below $T_{c}$, although (by construction) this divergence does occur in the pair propagators (\ref{pair-propagator-below_Tc}).
This strategy will enable us to connect with continuity with the results obtained in Ref.~\cite{Pisani-2018} for the normal phase \cite{footnote-6}. 
Our arguments go as follows.

The matrix elements of the inverse $T^{-1}$ of the matrix $T$ given by Eq.~(\ref{pair-propagator-below_Tc}) are:
\begin{equation}
\left( 
\begin{array}{cc}
T^{-1}(q)_{11}  &  T^{-1}(q)_{12}  \\
T^{-1}(q)_{21}  &  T^{-1}(q)_{22}
\end{array} 
\right)
= -
\left( 
\begin{array}{cc}
A(q)    &  B(q)  \\
B(q)  &    A(-q)
\end{array} 
\right)
\label{inverse-pair-propagator-below_Tc}
\end{equation}
\noindent
with the expressions (\ref{definition-A}) and (\ref{definition-B}) for $A(q)$ and $B(q)$.
The theory can be endowed by some degree of self-consistency, by replacing the above matrix $T^{-1}$ with a new matrix $\bar{T}^{-1}$ given by
\begin{eqnarray}
&& \left( \!
\begin{array}{cc}
\bar{T}^{-1}(q)_{11}  &  \bar{T}^{-1}(q)_{12}  \\
\bar{T}^{-1}(q)_{21}  &  \bar{T}^{-1}(q)_{22}
\end{array} 
\! \right)
\nonumber \\
& = &
- \left( \!\!
\begin{array}{cc}
A(q) + \Sigma^{\mathrm{B}}(0)_{11}  \,\,  &  B(q) + \Sigma^{\mathrm{B}}(0)_{12} \\
B(q) + \Sigma^{\mathrm{B}}(0)_{21}  \,\,  &  \!\!  A(-q) + \Sigma^{\mathrm{B}}(0)_{22}
\end{array} 
\!\! \right)
\label{inverse-pair-propagator-below_Tc}
\end{eqnarray}
\noindent
in terms of the constant shifts $\Sigma^{\mathrm{B}}(0)_{11}=\Sigma^{\mathrm{B}}(0)_{22}$ and $\Sigma^{\mathrm{B}}(0)_{12}=\Sigma^{\mathrm{B}}(0)_{21}$.
Here, $\Sigma^{\mathrm{B}}(0)_{11}$ refers either to the Popov term (\ref{Popov-bosonic-self-energy}) or to the GMB term (\ref{GMB-bosonic-self-energy_11}) or to the sum of both of them, while
$\Sigma^{\mathrm{B}}(0)_{12}$ refers to the GMB term (\ref{GMB-bosonic-self-energy_12}).
In all cases, the Hugenholtz-Pines condition for fermion pairs reads quite generally:
\begin{equation}
A(0) - B(0) + \Sigma^{\mathrm{B}}(0)_{11} - \Sigma^{\mathrm{B}}(0)_{12} = 0 
\label{generalized-HP-condition}
\end{equation}
\noindent
which recovers alternatively Eqs.~(\ref{Popov-gap-equation-II}) or (\ref{GMB-gap-equation-II}), depending on the choice of the bosonic-like self-energies $\Sigma^{\mathrm{B}}(0)_{ij}$. 
The condition (\ref{generalized-HP-condition}) guarantees, in addition, that the ``dressed'' pair propagator $\bar{T}$, too, is gapless at $q=0$.
With the form (\ref{inverse-pair-propagator-below_Tc}) of the matrix $\bar{T}^{-1}(q)$, one readily calculates its inverse $\bar{T}(q)$, whose matrix elements can, in turn, be introduced in the expressions 
(\ref{Popov-bosonic-self-energy}) of the Popov bosonic-like self-energy and/or (\ref{GMB-bosonic-self-energy_11}) and (\ref{GMB-bosonic-self-energy_12}) of the GMB bosonic-like 
self-energies, which are thus calculated with the dressed pair propagator $\bar{T}$ in the place of the bare $T$.

For the needs of the BCS-BEC crossover, the numerical solution of the Hugenholtz-Pines condition (\ref{generalized-HP-condition}) has to be determined in conjunction with that of the density equation 
(\ref{density-equation-t_matrix}), in which we keep $\mathcal{G}_{11}$ of the form (\ref{G_11-convenient_form}) and (\ref{sigma_11}).
Here, the (diagonal) fermionic self-energy $\Sigma_{11}$ has still the form (\ref{Sigma_11-t_matrix}), although with the matrix element $T_{11}(q)$ now replaced by $\bar{T}_{11}(q)$ 
so as to include a (partial) degree of self-consistency also in the density equation \cite{footnote-7}.
In this way, both the gap parameter $\Delta$ \emph{and\/} the chemical potential $\mu$ can be obtained for any given coupling $(k_{F} a_{F})^{-1}$ and temperature $T<T_{c}$, the process being
iterated until self-consistency is achieved.

The above procedure can be somewhat simplified, by exploiting the fact that the \emph{constant} shifts $\Sigma^{\mathrm{B}}(0)_{ij}$ in the matrix elements (\ref{inverse-pair-propagator-below_Tc}) enter also 
the Hugenholtz-Pines condition (\ref{generalized-HP-condition}).
Accordingly, with the use of Eq.~(\ref{generalized-HP-condition}) we can write for the diagonal elements in Eq.~(\ref{inverse-pair-propagator-below_Tc}):
\begin{equation}
A(\pm q) + \Sigma^{\mathrm{B}}(0)_{11} = A(\pm q) - A(0) + B(0) + \Sigma^{\mathrm{B}}(0)_{12} \, .
\label{A(q)-A0+B0+Sigma_12}
\end{equation}
\noindent
Here, we note that the difference
\begin{equation}
A(\pm q) - A(0) = \sum_{k} \left[ \mathcal{G}_{11}^{\mathrm{BCS}}(k \pm q) - \mathcal{G}_{11}^{\mathrm{BCS}}(k) \right] \mathcal{G}_{22}^{\mathrm{BCS}}(k) 
\label{A(q)-A0-explicit}
\end{equation}
\noindent
contains no \emph{explicit} reference to the coupling $(k_{F} a_{F})^{-1}$, which would otherwise enter the definition 
\begin{equation}
A(q) = - \frac{m}{4\pi a_{F}}  + \sum_{k} \mathcal{G}_{11}^{\mathrm{BCS}}(k+q) \, \mathcal{G}_{22}^{\mathrm{BCS}}(k) + \int \! \frac{d\mathbf{k}}{(2 \pi)^3} \, \frac{m}{\mathbf{k}^2}
\label{definition-A-regularized}
\end{equation}
\noindent
once $A(q)$ is suitably normalized in terms of the single-particle density of states $N_{0} = m k_{F}/(2 \pi^{2})$ per spin component.
[Note that the regularization condition (\ref{regularization}) has been used to obtain the expression (\ref{definition-A-regularized}) from the original definition (\ref{definition-A})).]
In addition, in this way the matrix elements (\ref{inverse-pair-propagator-below_Tc}) contain \emph{only\/} $\Sigma^{\mathrm{B}}(0)_{12}$, since $\Sigma^{\mathrm{B}}(0)_{11}$ has been eliminated therein through 
the use of the identity (\ref{A(q)-A0+B0+Sigma_12}). 

With these premises, it is convenient to organize the procedure of self-consistency in the following way:

\vspace{0.1cm}
\noindent
(i) Begin by fixing a pair of values $(T,\mu)$ which are expected to lie in the superfluid phase below $T_{c}$ (where the pair $(T_{c},\mu_{c})$ can be desumed from the results of Ref.~\cite{Pisani-2018}).

\vspace{0.1cm}
\noindent
(ii) Select further a value of $\Delta$ and calculate the quantities $A(\pm q) - A(0)$, $B(q)$, and $\Sigma^{\mathrm{B}}(0)_{12}$, to obtain the matrix elements (\ref{inverse-pair-propagator-below_Tc}) of 
$\bar{T}^{-1}(q)$ and of its inverse $\bar{T}(q)$.

\vspace{0.1cm}
\noindent
(iii) Enter the matrix element $\bar{T}(q)_{11}$ obtained in this way into the fermionic self-energy (\ref{Sigma_11-t_matrix}) in the place of $T(q)_{11}$, and use this self-energy to determine a new value of 
$\Delta$ which is consistent with the density equation (\ref{density-equation-t_matrix}).
 
\vspace{0.1cm} 
\noindent
(iv) Use this new value of $\Delta$ to calculate again the matrix elements of $\bar{T}^{-1}(q)$ with the help of Eqs.~(\ref{A(q)-A0+B0+Sigma_12}) and (\ref{A(q)-A0-explicit}), to be used once again in the density
equation to determine a new value $\Delta$. 
Repeat this process until self-consistency is achieved for $\Delta$.

\vspace{0.1cm}
\noindent
(v) Calculate $\Sigma^{\mathrm{B}}(0)_{11}$ (or, better, directly the difference $\Sigma^{\mathrm{B}}(0)_{11} - \Sigma^{\mathrm{B}}(0)_{12}$) with the values of $(T,\mu,\Delta)$ determined in this way.

\vspace{0.1cm}
\noindent
(vi) Insert $A(0)$ in the form (\ref{definition-A-regularized}), $B(0)$, and $\Sigma^{\mathrm{B}}(0)_{11} - \Sigma^{\mathrm{B}}(0)_{12}$ thus determined into the Hugenholt-Pines condition (\ref{generalized-HP-condition}), to obtain the corresponding value of the coupling $(k_{F} a_{F})^{-1}$.

The above procedure is somewhat more involved than that considered in Ref.~\cite{Pisani-2018} for the normal phase, where one was only interested in calculating $T_{c}$ (and the associated $\mu_{c}$) for given coupling.
In the superfluid phase of interest here, on the other hand, for given coupling one is required to determine the \emph{full\/} temperature dependence of $\Delta(T)$ and $\mu(T)$, from $T=0$ to $T_{c}$. 

Finally, we can also exploit the numerical procedures developed in Ref.~\cite{Pisani-2018} for the calculation of $\Sigma^{\mathrm{B}}(0)$ in the normal phase and utilise them now for the calculation of 
$\Sigma^{\mathrm{B}}(0)_{11}$ and $\Sigma^{\mathrm{B}}(0)_{12}$ in the superfluid phase.
To this end, it will be necessary to bring the expressions (\ref{Popov-bosonic-self-energy}) and (\ref{GMB-bosonic-self-energy_11}) for $\Sigma^{\mathrm{B}}(0)_{11}$ and (\ref{GMB-bosonic-self-energy_12}) for $\Sigma^{\mathrm{B}}(0)_{12}$ in the superfluid phase to a form that can be readily translated into that of $\Sigma^{\mathrm{B}}(0)$ in the normal phase.
As a consequence, the bosonic-like self-energies (\ref{Popov-bosonic-self-energy}), (\ref{GMB-bosonic-self-energy_11}), and  (\ref{GMB-bosonic-self-energy_12}) will be amenable to numerical computation essentially with the same level of effort encountered in Ref.~\cite{Pisani-2018} for the normal phase.
This strategy is discussed in detail in Appendix~\ref{sec:appendix-C}.

\begin{figure}[t]
\begin{center}
\includegraphics[width=8.0cm,angle=0]{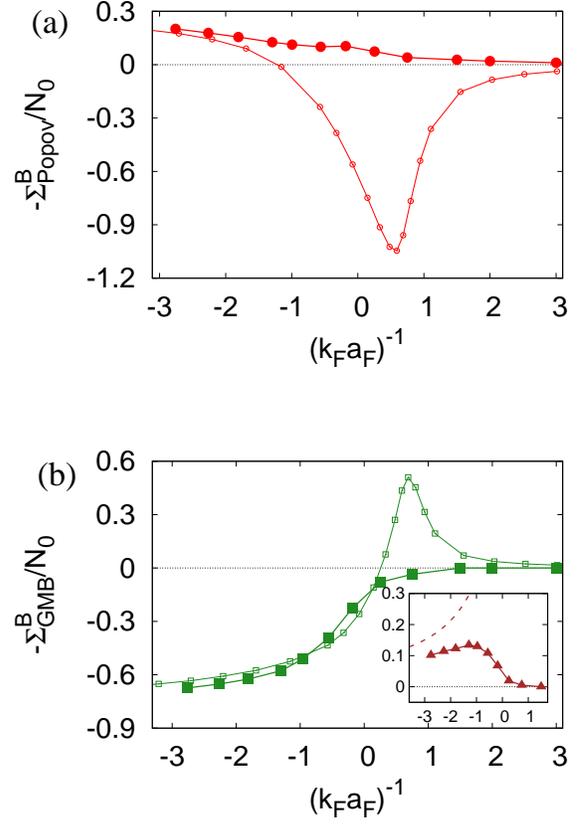}
\caption{(Color online) Bosonic-like self-energy $\Sigma^{\mathrm{B}}$ within (a) the Popov and (b) GMB approximations (multiplied by a minus sign and in units of the single-particle 
                                     density of states $N_{0} = m k_{F}/(2 \pi^{2})$ per spin component) vs the coupling $(k_{F} a_{F})^{-1}$. 
                                     Large (small) symbols refer to $T = 0$ ($T = T_{c}$).
                                     The inset in (b) shows the anomalous counterpart [-$\Sigma^{\mathrm{B}}_{12}(0)$] at $T=0$ within the GMB approximation (triangles) and its analytic 
                                     behaviour in the BCS regime (dashed line).}
\label{Figure-4}
\end{center} 
\end{figure}

\vspace{0.05cm}
\begin{center}
{\bf B. Bosonic-like self-energies that enter the Hugenholtz-Pines condition for fermion pairs}
\end{center}
\vspace{-0.2cm}

We have explicitly calculated numerically the bosonic-like self-energies $\Sigma^{\mathrm{B}}_{ij}$ within the Popov [cf. Eq.~(\ref{Popov-bosonic-self-energy})] and the GMB [cf. Eqs.~(\ref{GMB-bosonic-self-energy_11}) and (\ref{GMB-bosonic-self-energy_12})] approximations, for all temperatures below $T_{c}$ and couplings across the BCS-BEC crossover.
In both cases, we shall refer to the difference $\Sigma^{\mathrm{B}}_{11} - \Sigma^{\mathrm{B}}_{12}$ that enters the Hugenholtz-Pines condition (\ref{generalized-HP-condition}) for fermion pairs  
simply as $\Sigma^{\mathrm{B}}$ \cite{footnote-5}.

Figure~\ref{Figure-4} shows $\Sigma^{\mathrm{B}}$ (multiplied by a minus sign) throughout the BCS-BEC crossover and for the temperatures $T=0$ and $T=T_{c}$, within the Popov (upper panel) and GMB (lower panel) approximations.
For both temperatures, the Popov and GMB contributions to $\Sigma^{\mathrm{B}}$ have comparable magnitude over the whole coupling range, while the anomalous counterpart 
$\Sigma^{\mathrm{B}}_{12}(0)$ at $T=0$ (shown in the inset of panel (b) together with its analytic behaviour obtained from the expression (\ref{GMB-bosonic-self-energy_12-final-II}) 
in the BCS regime) turns out to be somewhat smaller.
In both Popov and GMB cases a marked difference appears between $T=0$ and $T_{c}$.
We have also verified (although not reported in the figure) that in both Popov and GMB cases a smooth evolution occurs as a function of temperature between the curves for $T=0$ 
and $T=T_{c}$ \cite{footnote-6}.  

\begin{figure}[t]
\begin{center}
\includegraphics[width=8.0cm,angle=0]{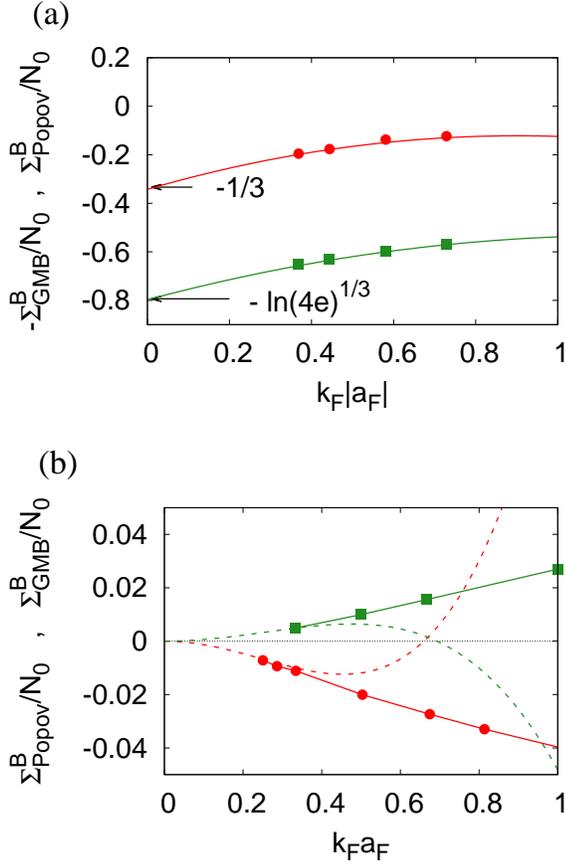}
\caption{(Color online)  (a) $\Sigma_{\mathrm{Popov}}^{\mathrm{B}}$ (circles) and $-\Sigma_{\mathrm{GMB}}^{\mathrm{B}}$ (squares) (in units of the single-particle 
                                          density of states $N_{0}$) vs $k_{F}|a_{F}|$ (with $a_{F} < 0$) obtained numerically at $T=0$ in the interval $(0,1)$. 
                                          The limiting values for $k_{F}|a_{F}| \rightarrow 0$ are seen to recover the respective analytic results.
                                     (b) $\Sigma_{\mathrm{Popov}}^{\mathrm{B}}$ (circles) and $\Sigma_{\mathrm{GMB}}^{\mathrm{B}}$ (squares) vs $k_{F}a_{F}$ (with $a_{F} > 0$) obtained numerically 
                                          at $T=0$ in the interval $(0,1)$ are compared with the respective analytic behaviours (dashed lines).
                                       In each case, the symbols are connected by a solid line obtained by a quadratic interpolation procedure.}
\label{Figure-5}
\end{center} 
\end{figure}

Figure~\ref{Figure-5} highlights the limiting behaviour of $\Sigma^{\mathrm{B}}$ within the Popov (circles) and GMB (squares) approximations
in the extreme BCS (upper panel) and BEC (lower panel) sides of the crossover.
In both panels, the lines have been drawn by a quadratic interpolation through the symbols.
In the extreme BCS regime, the limiting analytic values of $\Sigma_{\mathrm{Popov}}^{\mathrm{B}}/N_{0}$ $(=-1/3)$ and $\Sigma_{\mathrm{GMB}}^{\mathrm{B}}/N_{0}$ 
$(=\ln(4e)^{1/3})$ are seen to be accurately recovered by our numerical calculations.
In the extreme BEC regime, on the other hand, our numerical calculations are compared with the analytic expressions (dashed lines) reported in Appendix~\ref{sec:appendix-B}, where contributions
from both the non-condensed ($n'$) and condensed ($n_{0}$) densities are present.
Since at low temperature $n' \ll n_{0}$, the numerical effort to reach the extreme BEC regime in the superfluid phase is much more severe than in the normal phase where only $n' = n$ appears.

\vspace{0.05cm}
\begin{center}
{\bf C. Gap parameter throughout the BCS-BEC crossover}
\end{center}
\vspace{-0.2cm}

Once the quantity $\Sigma^{\mathrm{B}}$ is calculated numerically with due confidence and its analytic BCS and BEC limiting behaviours are suitably recovered, 
one can pass to determine the temperature and coupling dependence of the gap parameter $\Delta$ from the Hugenholtz-Pines condition (\ref{generalized-HP-condition}). 
This is done here within alternative approximations, namely, the Popov and GMB-plus-Popov approximations discussed in subsections~\ref{sec:G-MB-BCS-BEC}-C and \ref{sec:G-MB-BCS-BEC}-D 
(besides the standard mean-field approximation to compare with).

\begin{figure}[t]
\begin{center}
\includegraphics[width=8.7cm,angle=0]{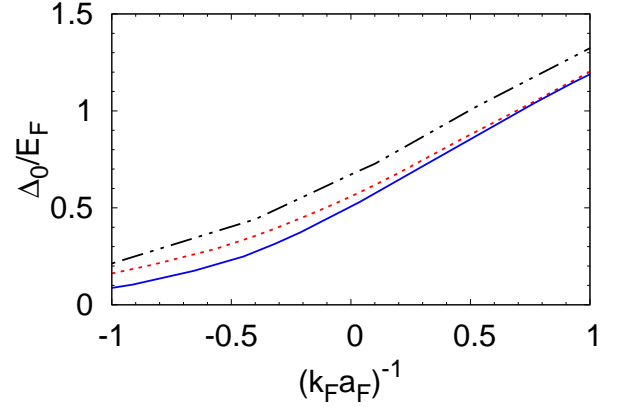}
\caption{(Color online) The gap parameter $\Delta_{0}$ at zero temperature (in units of the Fermi energy $E_{F}$) is shown vs the coupling $(k_{F}a_{F})^{-1}$ within three 
                                     different approximations: mean field (dashed double-dotted line); Popov (dashed line); GMB-plus-Popov (solid line).}
\label{Figure-6}
\end{center} 
\end{figure}

Figure~\ref{Figure-6} shows the coupling dependence of the gap parameter $\Delta_{0}$ at zero temperature, obtained within the above three approximations in the crossover region $-1 \lesssim (k_{F}a_{F})^{-1} \lesssim +1$ of most interest.
It is seen that the value of $\Delta_{0}$ systematically decreases over the whole coupling range, when passing from the mean-field, to the Popov, and then to the GMB-plus-Popov approximations,
where at each step higher degrees of pairing fluctuations beyond mean field are progressively taken into account.

\begin{figure}[t]
\begin{center}
\includegraphics[width=7.5cm,angle=0]{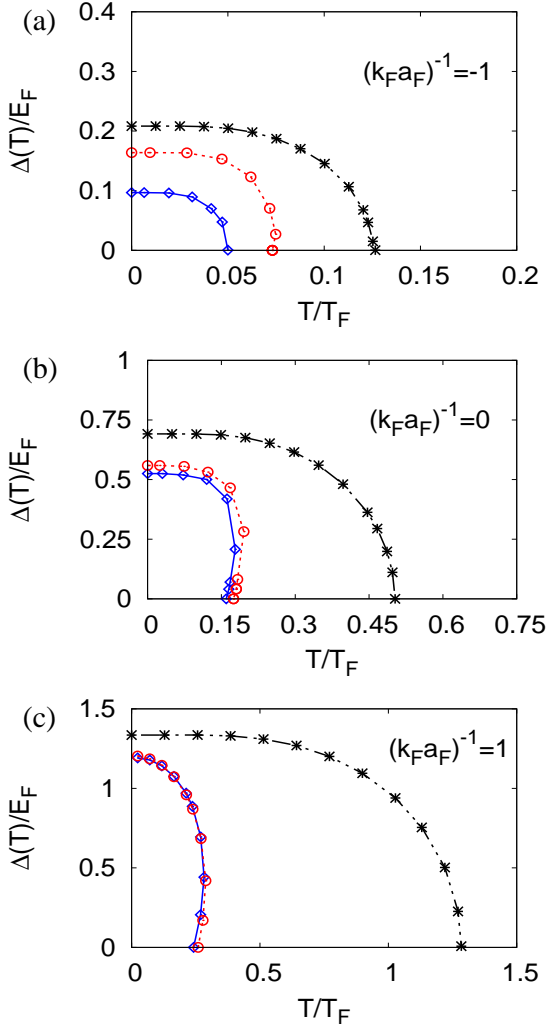}
\caption{(Color online) Temperature dependence of the gap parameter $\Delta(T)$ (in units of the Fermi energy $E_{F}$) for the couplings 
                                    (a) $(k_{F} a_{F})^{-1} = -1.0$, (b) $(k_{F} a_{F})^{-1} = 0.0$ and (c) $(k_{F} a_{F})^{-1} = +1.0$, 
                                     within the mean-field (stars), Popov (circles), and GMB-plus-Popov (diamonds) approximations.}
\label{Figure-7}
\end{center} 
\end{figure}

\begin{figure}[t]
\begin{center}
\includegraphics[width=7.2cm,angle=0]{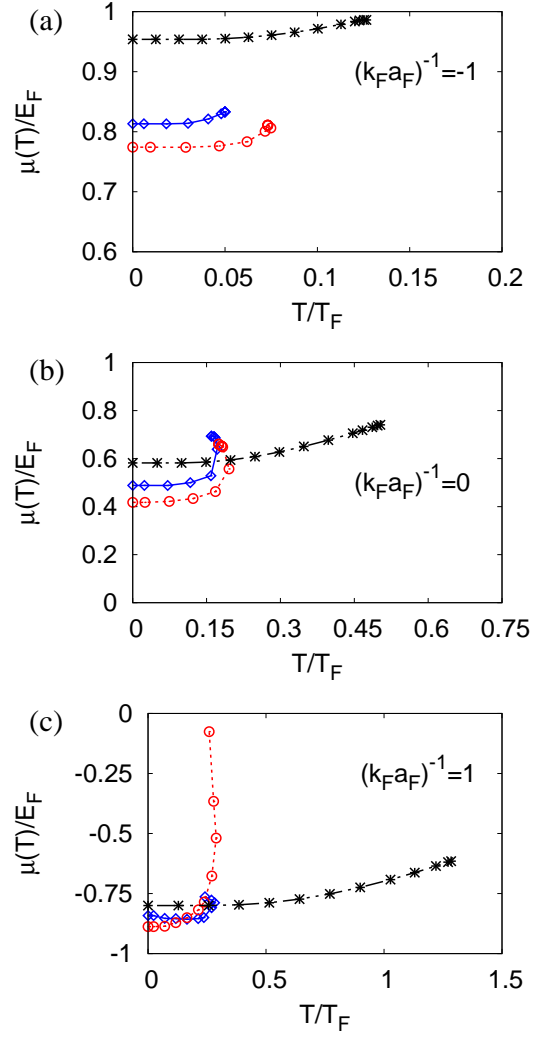}
\caption{(Color online) Temperature dependence of the chemical potential $\mu(T)$ (in units of the Fermi energy $E_{F}$) for the couplings 
                                     (a) $(k_{F} a_{F})^{-1} = -1.0$, (b) $(k_{F} a_{F})^{-1} = 0.0$, and (c) $(k_{F} a_{F})^{-1} = +1.0$, 
                                     within the mean-field (stars), Popov (circles), and GMB-plus-Popov (diamonds) approximations.}
\label{Figure-8}
\end{center} 
\end{figure}

The temperature dependence $\Delta(T)$ of the gap parameter is reported in Fig.~\ref{Figure-7} within the above three approximations and for three characteristic couplings.
Several interesting features can be highlighted from these plots, while comparing, in particular, the results of the GMB-plus-Popov approximation with those of mean field.
When including pairing fluctuations beyond mean field, the suppression of the gap $\Delta_{0}$ at $T=0$ is less pronounced than the corresponding reduction of the critical temperature $T_{c}$.
For instance, at unitarity $\Delta_{0} = 0.525 E_{F}$ and $T_{c} = 0.160 E_{F}$ within the GMB-plus-Popov approximation, such that $\Delta_{0}/T_{c} = 3.281$;
conversely, within mean field $\Delta_{0} = 0.687 E_{F}$ and $T_{c} = 0.50 E_{F}$, such that $\Delta_{0}/T_{c} = 1.339$ (a value smaller than the result $1.76$ valid in the extreme BCS limit
$(k_{F} a_{F})^{-1} \ll -1$,  also once the GMB contribution is included \cite{GMB-1961}).
As a consequence, the curve $\Delta(T)$ within the GMB-plus-Popov approximation gets somewhat more compressed along the $T$-axis than along the $\Delta$-axis, when compared with the corresponding
curve obtained within mean field.
This feature appears evident in all three panels of Fig.~\ref{Figure-7}.
Owing to this nonuniform compression of the curve, when including pairing fluctuations beyond mean field $\Delta(T)$ remains closer to its zero-temperature value $\Delta_{0}$ over 
a wider portion of the temperature interval up to $T_{c}$ when compared with mean field, and then falls rather abruptly to zero only quite close to $T_{c}$.
This behaviour is reminiscent of what found experimentally for the temperature dependence of the superfluid fraction in a ultra-cold Fermi gas \cite{Grimm-2013}, 
which remains almost completely superfluid below $0.6T{c}$.
Finally, a comment is in order about the ``reentrant'' behaviour found for $\Delta(T)$ when $T$ approaches $T_{c}$, which develops gradually when passing from the BCS to the BEC side of the crossover as seen in Fig.~\ref{Figure-7} (although this is less evident in the GMB-plus-Popov than in the Popov approximation).
This behaviour is inherited from the Bogoliubov-Popov theory for point-like bosons, to which the condensate density presents a similar behaviour \cite{Luban-1966,Luban-1970,Shi-1998} and to which the present theory reduces in the BEC limit of tightly-bound composite bosons (although a minor reentrant behaviour begins to show up in the crossover region where composite bosons are not yet fully developed).

Figure~\ref{Figure-8} shows related plots for the temperature dependence of the chemical potential.
This is seen to decrease monotonically below $T_{c}$, in line with the progressive building up of the condensate upon lowering the temperature (also for this quantity, the reentrant behaviour close to $T_{c}$ becomes  less evident when passing from the Popov to the GMB-plus-Popov approximation).

\begin{figure}[t]
\begin{center}
\includegraphics[width=7.1cm,angle=0]{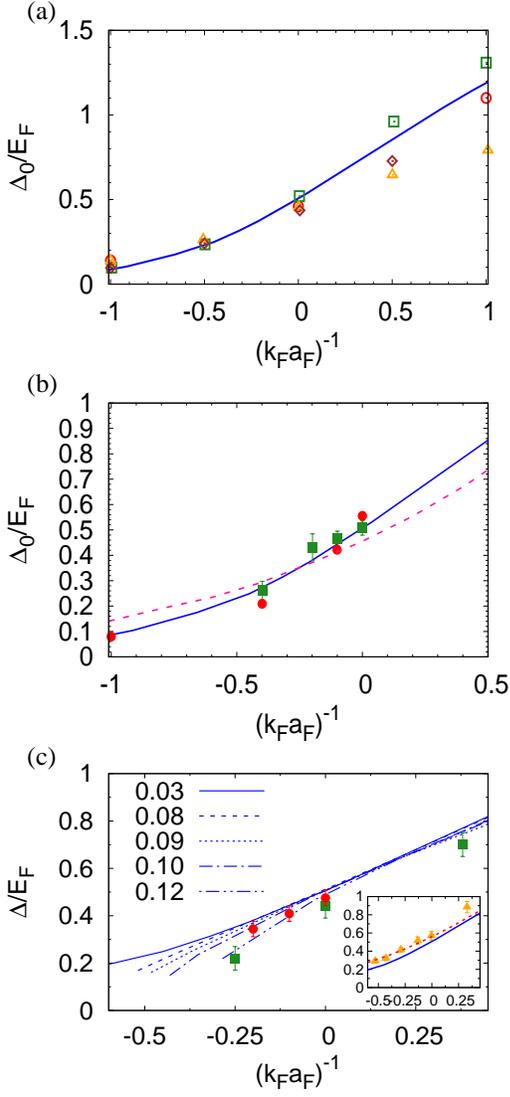}
\caption{(Color online) The results of the GMB-plus-Popov calculation for the gap parameter are compared with those obtained $T=0$ by (a) alternative diagrammatic or functional-integral 
                                     approaches and (b) QMC methods. In addition, panel (b) shows the theoretical results obtained by the diagrammatic approach of Ref.~\cite{Haussmann-2009} (broken line)
                                     In panel (c) the results of the GMB-plus-Popov calculation at finite temperature are compared with two independent sets of experimental data from Refs.~\cite{Ketterle-2008} and \cite{Vale-2017}. 
                                     The inset in panel (c) compares the GMB-plus-Popov (full line) and Popov (broken line) calculations with the experimental data from Ref.~\cite{Koehl-2018}.
                                     [The meaning of the lines and symbols (as well as the references from which the data are taken) is given in the text.]}
\label{Figure-9}
\end{center} 
\end{figure}

It is relevant to compare our results with those obtained by other theoretical (diagrammatic, functional-integral, and QMC) approaches, as well as with the available experimental data.
This comparison is shown in Fig.~\ref{Figure-9}.
Specifically, the coupling dependence of the zero-temperature gap $\Delta_{0}$ obtained by the present GMB-plus-Popov calculation (solid line) is compared in Fig.~\ref{Figure-9}(a)
with the results of the diagrammatic or functional-integral approaches of Refs.~\cite{Randeria-2008} (triangles), \cite{Haussmann-2009} (circles), \cite{Chen-2016} (squares), and \cite{Ohashi-2017} (diamonds),
and in Fig.~\ref{Figure-9}(b) with the QMC data from Refs.~\cite{Carlson-2008-b} (squares with error bars) and \cite{Bulgac-2008} (circles with error bars).
In addition, Fig.~\ref{Figure-9}(c) compares the experimental data from Refs.~\cite{Ketterle-2008} (squares with error bars) and \cite{Vale-2017} (circles with error bars), 
taken at low but non-zero temperatures, with our GMB-plus-Popov results calculated at $T=0$ (solid line), $T=0.08 T_{F}$ (dashed line), $T=0.09 T_{F}$ (dotted line), $T=0.10 T_{F}$ (dashed-dotted line), 
and $T=0.12 T_{F}$ (dashed double-dotted line).

In Fig.~\ref{Figure-9}(b) it is worth pointing out that to the present GMB-plus-Popov calculation (full line) there corresponds a steeper dependence on coupling about unitarity as
compared with the diagrammatic calculation of Ref.~\cite{Haussmann-2009} (broken line).
This steeper dependence, which is seen to reproduce the trend of the QMC data, is consistent with the stronger suppression of the gap on the BCS side of unitarity due to the GMB contribution.
We have verified that the presence of the anomalous bosonic-like self-energy $\Sigma_{\mathrm{GMB}}^{\mathrm{B}}(0)_{12}$ in the Hugenholtz-Pines condition (\ref{Popov_plus_GMB-HP}), 
which is a distinctive feature of the present GMB-plus-Popov calculation, contributes significantly to this steeper dependence, since it affects the value of $\Delta_{0}$ up to about $20 \%$ 
on the weak-coupling side of the crossover.

Particularly encouraging appears the comparison shown in Fig.~\ref{Figure-9}(c) between the experimental data and our results, just taken at the temperatures that correspond 
to those reported experimentally.
For instance, at unitarity Ref.~\cite{Vale-2017} gives the value $\Delta/E_{F} = 0.47 \pm 0.03$ for the temperature range $T/T_{F} = 0.09 \pm 0.01$.
Correspondingly, the GMB-plus-Popov calculation at unitarity yields $\Delta/E_{F} = (0.521,0.507,0.504,0.489)$ for the temperatures $T/T_{F} = (0.08,0.09,0.10,0.12)$, in the order.
This comparison also demonstrates that the effect of temperature acquires a growing importance for the gap as soon as one moves from the BEC into the BCS regime. 

In addition, the inset of Fig.~\ref{Figure-9}(c) compares the results of the GMB-plus-Popov (full line) and Popov (broken line) calculations at $T=0$ with the experimental data from Ref.~\cite{Koehl-2018} (triangles),
which are taken at the nominal temperature $T/T_{F} = 0.07 \pm 0.02$.
This set of experimental data appears to agree quite well with the Popov calculation, while discrepancies appear when compared with the GMB-plus-Popov calculation.
On the contrary, we have already commented that the GMB-plus-Popov calculation agree quite well with the experimental data of Ref.~\cite{Vale-2017}.
The difference between the two sets of experimental data could possibly be attributed to the different protocols adopted by the two experiments to extract the gap.
While Ref.~\cite{Vale-2017} measures a response (density-density correlation) function in the linear regime for which the system is probed at thermodynamic equilibrium, Ref.~\cite{Koehl-2018} adopts a
time-dependent protocol that brings the system out of thermodynamic equilibrium. 
This may give rise to a retardation mechanism, whereby increasingly complicated many body-processes (like the GMB contribution) could take longer time than simpler processes (like the Popov one) before being excited by the experimental protocol, in analogy to what occurs in the context of the orthogonality catastrophe \cite{Mahan-2000}.

\begin{figure}[t]
\begin{center}
\includegraphics[width=8.5cm,angle=0]{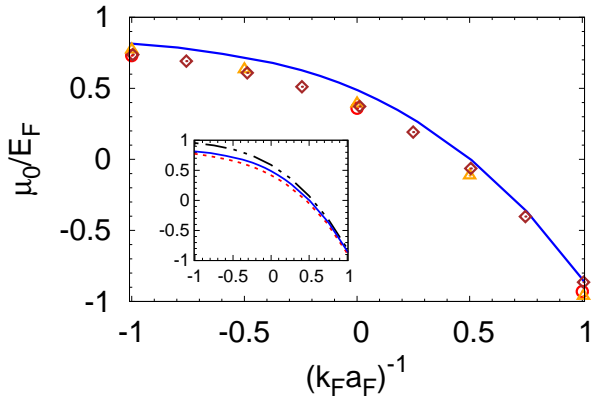}
\caption{(Color online) The coupling dependence of the chemical potential at zero temperature (in units of the Fermi energy $E_{F}$) obtained by the GMB-plus-Popov calculation (solid line) 
                                     is compared with the results by the diagrammatic or functional-integral approaches of Refs.~\cite{Randeria-2008} (triangles), \cite{Haussmann-2009} (circles), 
                                     and \cite{Ohashi-2017} (diamonds).
                                     The inset compares the results of the GMB-plus-Popov calculation (solid line) with those of the Popov (dashed line) and mean field (dashed double-dotted line) calculations.}
\label{Figure-10}
\end{center} 
\end{figure}

Finally, the coupling dependence of the chemical potential at zero temperature is shown in Fig.~\ref{Figure-10} for all the three (mean field, Popov, and GMB-plus-Popov) 
approximations considered in the present paper.
Our results are further compared with those obtained by alternative diagrammatic or functional-integral approaches.
The comparison shows that the GMB-plus-Popov results are systematically larger than those obtained by other approaches over the entire crossover region.
This outcome is in line what was that found in Ref.~\cite{Pisani-2018} when approaching $T_{c}$ from the normal phase.
There it was argued that, endowing the single-particle fermionic propagators that enter the expressions of the Popov and GMB bosonic-like self-energies with a suitable fermionic self-energy insertion, 
acts to decrease the values of the chemical potential without affecting at the same time the values of other thermodynamic quantities.
Translating this argument to the superfluid phase, we expect this conclusion to imply that also the gap parameter will not be affected by modifying the chemical potential along the above lines.

\vspace{0.05cm}
\begin{center}
{\bf D. Further analysis of the GMB contribution to the gap parameter}
\end{center}
\vspace{-0.2cm}

We conclude this Section by digging somewhat further on the GMB contribution to the gap parameter.

It first appears relevant to check how the numerical accuracy on the calculation of $\Sigma^{B}_{\mathrm{GMB}}$ (as well as of $\Sigma^{B}_{\mathrm{Popov}}$), that was considered in 
Fig.~\ref{Figure-5}, translates into the accuracy on the calculation of the gap parameter itself.
This check is particularly relevant in the extreme BCS limit $(k_{F} a_{F})^{-1} \ll -1$, for which the result (\ref{GMB-Delta-0}) at zero temperature was obtained analytically long ago by the original GMB work \cite{GMB-1961}, but it has never been recovered since through an accurate numerical calculation which would approach this limit from \emph{finite} values of $(k_{F} a_{F})^{-1}$.
\begin{figure}[t]
\begin{center}
\includegraphics[width=8.5cm,angle=0]{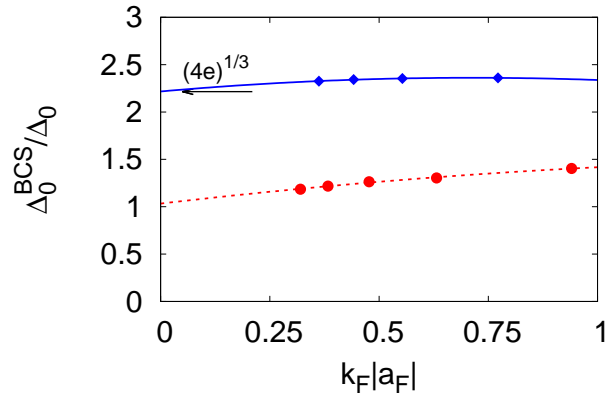}
\caption{(Color online) The ratio $\Delta_{0}^{\mathrm{BCS}}/\Delta_{0}$ at $T=0$ is shown in the BCS (weak-coupling) regime $k_{F}|a_{F}| \lesssim 1.0$ (with $a_{F}<0$), 
                                     within the GMB-plus-Popov (diamonds) and Popov (circles) approximations. 
                                     Here, $\Delta_{0}^{\mathrm{BCS}}$ is given by the mean-field expression (\ref{final-expression-gap-BCS-approximation-T=0-lower}) that holds in the BCS regime,
                                     and the lines represent quadratic interpolations through the symbols.}
\label{Figure-11}
\end{center} 
\end{figure}
This check is shown in Fig.~\ref{Figure-11}, where the values of $\Delta_{0}$ obtained at zero temperature within the the GMB-plus-Popov and Popov approximations 
over an extended range of (inverse) coupling $k_{F}|a_{F}|$ (which contains the extreme BCS limit $k_{F}|a_{F}| =0$) are compared with the mean-field values given by the expression 
(\ref{final-expression-gap-BCS-approximation-T=0-lower}). 
In the figure, the numerical results (diamonds and circles) are supplemented by a quadratic interpolation both for the GMB-plus-Popov (full line) and the Popov case (dashed line).
In both cases, these interpolations converge with extremely good accuracy to the expected value at $k_{F}|a_{F}| =0$ when the limit $k_{F}|a_{F}| \rightarrow 0$ is taken.

\begin{figure}[t]
\begin{center}
\includegraphics[width=8.5cm,angle=0]{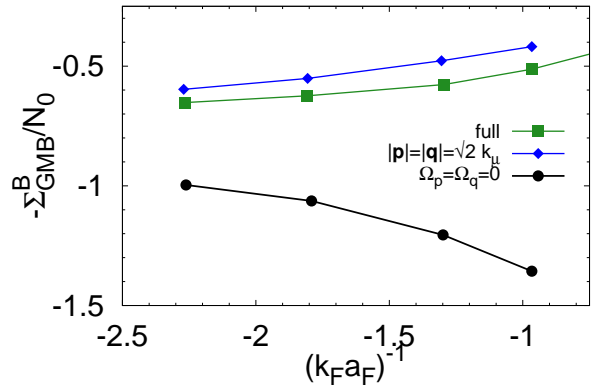}
\caption{(Color online) The full calculations of $-\Sigma_{\mathrm{GMB}}^{\mathrm{B}}$ vs $(k_{F}a_{F})^{-1}$ at  $T = 0$ is compared with partial calculations 
                                    of the same quantity, in which either the wave-vector or the frequency dependence has been neglected in both pair propagators $\bar{T}_{11}$ entering the expression of 
                                    $\Sigma_{\mathrm{GMB}}^{\mathrm{B}}$. 
                                    The corresponding values of $\mu$ obtained in the full calculation are also used in the partial calculations.}
\label{Figure-12}
\end{center} 
\end{figure}

An additional piece of information which is worth supplying is to what extent the wave-vector and frequency dependence of the pair propagator $T_{11}$ affects the numerical value of the 
GMB bosonic-like self-energy $\Sigma_{\mathrm{GMB}}^{\mathrm{B}}$ in the broken-symmetry phase, over an extended coupling range away from the extreme BCS limit.
This information is relevant in the present context, because the effects of the wave-vector and frequency dependence of the pair propagator on the GMB correction in the broken-symmetry phase 
were never considered before.
Although neglecting the wave-vector and frequency dependence of the pair propagator can be justified in the extreme BCS limit (as it was shown analytically in subsection~\ref{sec:G-MB-BCS-BEC}-D), 
we have emphasized throughout this paper that this cannot be the case when departing from the extreme BCS limit and spanning the BCS-BEC crossover.
As a further step, we can determine which one of the two dependences of the pair propagator $T_{11}$, namely, \emph{either} on the wave vector \emph{or} on frequency, turns out to be 
most important for $\Sigma_{\mathrm{GMB}}^{\mathrm{B}}$.
The result of this test is reported in Fig.~\ref{Figure-12}, where the calculation of $\Sigma_{\mathrm{GMB}}^{\mathrm{B}}$ at $T = 0$ with the full wave-vector and frequency dependence of $T_{11}$ is compared over an extended coupling range on the BCS side of unitarity with two partial calculations, in which $T_{11}$ has been deprived of either its wave-vector or frequency dependence.
This result shows that the frequency dependence of $T_{11}$ is by far the most dominant one for $\Sigma_{\mathrm{GMB}}^{\mathrm{B}}$, thus extending to the broken-symmetry phase an analogous result obtained in
Ref.~\cite{Pisani-2018} for the normal phase.

\section{Concluding remarks and perspectives}
\label{sec:conclusions}
\vspace{-0.3cm}

The motivation behind the work presented in this paper has been twofold.

The first aspect has been, quite generally, to cast the gap equation for superfluid fermions in an alternative and physically more appealing form, which would emphasize at the outset the composite nature of the fermion pairs, in such a way to make it more direct (and, possibly, more straightforward) the inclusion of pairing fluctuations beyond mean field in the gap equation itself.
In this context, we have proved that the gap equation is equivalent to a Hugenholtz-Pines condition for fermion pairs, which contains (normal and anomalous) bosonic-like self-energies that dress the bare pair propagator, in analogy to the original Hugenholtz-Pines condition for point-like bosons.  
The proof rests on the use of many-body diagrammatic methods and holds for any choice of a conserving (or, more generally,  just self-consistent) approximation for the fermionic self-energy that describes the constituent fermions 
(with the provision of always including in this choice at least the Fock-like term which is at the basis of the BCS theory of superconductivity).
To prove this equivalence, unnecessary details of the inter-particle interaction were eliminated by restricting to a contact interaction.

The second aspect has been to test (and, at the same time, to take direct advantage of) this new formulation for the gap equation, to address the long-pending problem of including the GMB correction to the gap equation in a systematic way.
And this not only in the BCS limit at zero temperature, as it was done in the original GMB work of Ref.~\cite{GMB-1961}, but also at any temperature in the superfluid phase below $T_{c}$ as well as across the whole 
BCS-BEC crossover.
At present, the need to span this crossover stems from the fact that experimental data with ultra-cold Fermi gases and QMC calculations have recently become available for the superfluid phase of a Fermi gas in the intermediate-coupling regime between the BCS and BEC regimes, thereby providing us with the opportunity to compare these data with the results of our diagrammatic calculations.
In this respect, the quite good agreement that has resulted, between our diagrammatic calculations that include the GMB correction and the experimental and QMC data (and this not only over an extended range of coupling but also as far as the temperature dependence is concerned), has rewarded us for the considerable numerical efforts required to bring these calculations to completion.

It is worth making a few final comments about the meaning which is attributed to the (superconducting/superfluid) gap $\Delta$ in related contexts.
In the original BCS theory of superconductivity \cite{BCS-1957,Schrieffer-1964}, the energy gap $\Delta$ had initially played the role of a ``thermodynamic'' \emph{parameter}, to be eliminated in favor of the thermodynamic variables (temperature and chemical potential or density) to minimize the grand-canonical thermodynamic potential within a mean-field decoupling.
Only at a later stage the \emph{same} quantity $\Delta$ was also interpreted as a ``dynamic'' pairing gap, inasmuch as it enters the energy dispersion $E_{\mathbf{k}}$ that appears in the fermionic propagators 
(\ref{BCS-propagator-11}) and (\ref{BCS-propagator-12}) (whereby a window of unaccessible states opens up in the single-particle density of states).
In this way, the value of $\Delta$ or $2 \Delta$ (with its associated temperature dependence) can be related to what is measured experimentally through single- or two-particle properties \cite{Tinkham-1980}.
However, this equivalence between thermodynamic and dynamic gap is, in principle, lost when pairing fluctuations beyond mean field are included.
This inclusion can be done either by diagrammatic or by functional-integral approaches \cite{AS-2010}, whereby in both cases the thermodynamic energy gap $\Delta$ is regarded as a parameter of the theory to be self-consistently determined, being directly related to the non-vanishing of the pair amplitude $\langle \psi_{\uparrow}(\mathbf{r}) \psi_{\downarrow}(\mathbf{r}) \rangle$ in the broken-symmetry phase.
As a consequence, direct access to the value of the dynamic pairing gap would require one to perform additional calculations, in order to determine the spectrum of the dynamic response function that corresponds to a given experimental set up.
Specifically, the dynamic pairing gap $\Delta$ was experimentally determined, in Ref.~\cite{Ketterle-2008} by examining radio-frequency spectra obtained with an imbalanced ultra-cold Fermi gas,
while in Ref.~\cite{Vale-2017} two-photon Bragg spectroscopy on a balanced ultra-cold Fermi gas gave access to $2 \Delta$.
In both cases, however, extracting $\Delta$ from the data has relied on a mean-field-like interpretation for the role played by $\Delta$ as a single-particle energy gap.
The QMC calculations mentioned in subsection \ref{sec:numerical-results}-C, on the other hand, determine the pairing gap $\Delta$ as a single-particle property, either directly by calculating the difference of the ground-state energies when the total number of particles is changed by one unit \cite{Carlson-2008-b}, or by fitting the profile of the single-particle spectral function with a BCS-like form for the quasi-particle dispersion 
$E_{\mathbf{k}}$ \cite{Bulgac-2008}.

In this context, it appears relevant the transmuting that was made by the present diagrammatic approach, of the equation for the (thermodynamic) gap parameter $\Delta$ into a Hugenholtz-Pines condition for fermion pairs. 
This is because, in this way, the gap equation itself was endowed with a dynamical character, to the extent that the Hugenholtz-Pines condition guarantees the dynamical Goldstone mode built up on fermion pairs to be gapless.
The same value of $\Delta$ that makes this possible should then also enter other excitations of the condensate, like single-particle pair-breaking excitations (related to $\Delta$) and the Higgs mode (related to $2 \Delta$).
It is thus relevant that the experiment carried out in Ref.~\cite{Vale-2017} (whose results for $\Delta$ we have extensively compared with) was able to determine \emph{simultaneously\/} both the Goldstone mode and
the pair-breaking excitations.
In addition, it appears that an accurate determination of the Higgs mode is nowadays feasible for an ultra-cold Fermi gas spanning the BCS-BEC crossover \cite{Koehl-2018} (see also Ref.~\cite{Harrison-2017}).
Future work along these lines should thus apply the present treatment of the GMB contribution to the gap parameter (possibly extended also at finite frequency), to determine how it would affect the mixing between the Goldstone and Higgs modes while evolving along the BCS-BEC crossover.


\begin{center}
\begin{small}
{\bf ACKNOWLEDGMENTS}
\end{small}
\end{center}
\vspace{-0.1cm}

We are indebted to C. J. Vale for a discussion about the comparison between the experimental data of Ref.~\cite{Vale-2017} and our theoretical calculations, and to R. Haussmann and W. Zwerger 
for providing us with the numerical results of the self-consistent $t$-matrix approximation reported in Fig.\ref{Figure-9}.
This work was partially supported by the Italian MIUR under Contract PRIN-2015 No. 2015C5SEJJ001.

\appendix   
\section{SUMMARY OF THE T-MATRIX APPROXIMATION BELOW $T_{c}$}
\label{sec:appendix-A}
\vspace{-0.2cm}

In this Appendix, we briefly recall the main features of the $t$-matrix approximation in the broken-symmetry phase \cite{Andrenacci-2003}, which are systematically used in this paper.

Quite generally, the \emph{many-particle T-matrix} is defined as the solution to the equation
\begin{eqnarray}
T(1,2;1',2') & = & \Xi(1,2;1',2') + \int \! d3456 \, \Xi(1,4;1',3)
\nonumber \\
& \times & \mathcal{G}(3,6) \, \mathcal{G}(5,4) \, T(6,2;5,2')
\label{definition-many-particle-T-matrix}
\end{eqnarray}
\noindent
where
\begin{equation}
\Xi(1,2;1',2') = \frac{\delta \Sigma(1,1')}{\delta \mathcal{G}(2',2)}
\label{definition-effective-two-particle-interaction}
\end{equation}
\noindent
is the \emph{effective two-particle interaction} obtained by functional differentiation of the fermionic self-energy $\Sigma$ with respect to the single-particle fermionic propagator $\mathcal{G}$.
The indices $1, 2, \cdots$ are a shorthand notation for the spatial coordinate $\mathbf{r}$, imaginary time $\tau$, and Nambu index $\ell$.

In particular, to the Fock-like term of Fig.~\ref{Figure-1}(b) there corresponds the expression
\begin{equation}
\Sigma_{\mathrm{Fock}}(1,1') = - V(x_{1}^{+} - x_{1'}) \sum_{\ell,\ell'=1}^{2} \tau^{3}_{\ell_{1}\ell} \mathcal{G}(x_{1},x_{1'})_{\ell,\ell'} \tau^{3}_{\ell'\ell_{1'}} 
\label{Fock-like-term}
\end{equation}
\noindent
with $\ell_{1}=1$ and $\ell_{1'}=2$ for the anomalous component we are interested in.
In the expression (\ref{Fock-like-term}), $\tau^{3}$ is a Pauli matrix, the Nambu indices have been made explicit by setting $x=(\mathbf{r},\tau)$ such that $1=(x_{1},\ell_{1})$ and so on, and 
the inter-particle interaction has been generically indicated by $V(x-x')$
(although we shall take eventually $V(x-x') = v_{0} \delta(\mathbf{r}-\mathbf{r'}) \delta(\tau-\tau')$  with $v_{0}<0$).

To the choice (\ref{Fock-like-term}) of the anomalous self-energy there corresponds an effective two-particle interaction of the form:
\begin{eqnarray}
& & \Xi_{\mathrm{Fock}}(1,2;1',2') = \frac{\delta \Sigma_{\mathrm{Fock}}(1,1')}{\delta \mathcal{G}(2',2)} = -  (1-\delta_{\ell_{1}\ell_{1'}}) 
\nonumber \\
& \times & \tau^{3}_{\ell_{1}\ell_{2'}} \delta(x_{1} - x_{2'}) V(x_{1}^{+}-x_{1'}) \delta(x_{1'} - x_{2}) \tau^{3}_{\ell_{1'}\ell_{2}} 
\label{effective-two-particle-interaction-Fock}  
\end{eqnarray}
where we have remarked that $\ell_{1} \neq \ell_{1'}$ consistently with the choice (\ref{Fock-like-term}).
Owing to this restriction, only four elements of the many-particle T-matrix survive in Nambu space, namely, those with $\ell_{1} \neq \ell_{1'}$ and $\ell_{2} \neq \ell_{2'}$.
Following Ref.~\cite{Andrenacci-2003}, it is then convenient to adopt the short-hand convention $1 \leftrightarrow (\ell=1,\ell'=2)$ and $2 \leftrightarrow (\ell=2,\ell'=1)$ to label the non-vanishing matrix elements of the T-matrix.
Upon Fourier transforming from $x$- to $q$-space, the matrix elements of the ($2 \times 2$) T-matrix are eventually given by the expression (\ref{general-pair-propagator-below_Tc}) for the case of 
interest of a contact inter-particle interaction.

\section{POPOV AND GMB BOSONIC-LIKE SELF-ENERGIES BELOW $T_{c}$ IN THE BEC LIMIT}
\label{sec:appendix-B}
\vspace{-0.2cm}

In this Appendix, the BEC (strong-coupling) limit of the Popov and GMB bosonic-like self-energies in the broken-symmetry phase is considered in detail.
It is shown that each of these structures contains two distinct diagrammatic contributions to the scattering length for composite bosons, which are made up of tight fermion pairs.
These two contributions originate from the distinct $q$-behaviors of the normal component $T_{11}(q)$ of the $t$-matrix in the broken-symmetry phase, which are relevant, respectively, over
the bosonic ($\mu_{B}$) and fermionic ($\mu$) energy scales, where $\mu_{B} = 2\mu + \epsilon_{0}$ with $\epsilon_{0} =  (m a_{F}^{2})^{-1}$ the binding energy of the two-fermion problem.
Specifically, over the bosonic energy scale $\mu_{B}$, $T_{11}(q)$ acquires the Bogoliubov form:
\begin{equation}
T_{11}(q) = - \frac{8 \pi}{m^{2} a_{F}} \frac{ \frac{\mathbf{q}^{2}}{4m} + i \Omega_{\nu} +\mu_{B}}{E_{B}(\mathbf{q})^{2} - (i \Omega_{\nu})^{2}}
\label{bosonic-expression-T_11}
\end{equation}
\noindent
where $E_{B}(\mathbf{q}) = \sqrt{\left( \frac{\mathbf{q}^{2}}{4m} + \mu_{B} \right)^{2} - \mu_{B}^{2}}$ is the dispersion relation for composite bosons \cite{Andrenacci-2003}.
\noindent
Over the fermionic energy scale $\mu$, on the other hand, $T_{11}(q)$ reduces to the asymptotic form \cite{Pisani-2004}:
\begin{equation}
T_{11}(q) = \frac{1}{ \frac{m}{4 \pi a_{F}} - \frac{m^{3/2}}{4 \pi} \sqrt{\frac{\mathbf{q}^{2}}{4m} - i \Omega_{\nu} - 2 \mu}} \, .
\label{asymptotic-expression-T_11}
\end{equation}
\noindent
It turns out that, in the diagrammatic expressions of the Popov and GMB bosonic-like self-energies, the above two distinct contributions to $T_{11}(q)$ are alternatively picked up, in the order,
depending on whether $T_{11}(q)$ is summed over $q$ together with a companion single-particle fermionic propagator of the type $\mathcal{G}_{0}(q)$ or $\mathcal{G}_{0}(-q)$. 
In the latter case, the diagram of interest contains a sub-unit corresponding to a scattering process that contributes to the scattering length $a_{B}$ for composite bosons \cite{Kagan-2005} 
(over and above the Born contribution), as it will be explicitly confirmed by the examples below.

\vspace{0.05cm}
\begin{center}
{\bf A. Popov contribution below $T_{c}$ in the BEC limit}
\end{center}
\vspace{-0.2cm} 

The Popov bosonic-like self-energy in the broken-symmetry phase is given by the expression (\ref{Popov-bosonic-self-energy}), where
only the fermionic propagator $\mathcal{G}_{22}^{\mathrm{BCS}}(k-q')$ entangles with the element $T_{11}(q')$ of the T-matrix.
With the help of the approximate expansion (\ref{approximate-BCS-propagator-11}) valid in the BEC limit, two terms are seen to contribute to the right-hand side of the expression (\ref{Popov-bosonic-self-energy}).
We thus write:
\begin{equation}
\Sigma_{\mathrm{Popov}}^{\mathrm{B}} = \Sigma_{\mathrm{Popov}}^{\mathrm{B}}(\mathrm{I}) + \Sigma_{\mathrm{Popov}}^{\mathrm{B}}(\mathrm{II}) \, .
\label{two-terms-Popov}
\end{equation}
\noindent
The first term in Eq.~(\ref{two-terms-Popov}) reads:
\begin{small}
\begin{eqnarray}
\Sigma_{\mathrm{Popov}}^{\mathrm{B}}(\mathrm{I}) & = &
- 2 \sum_{k,q}  \mathcal{G}_{0}(k)^{2} \, \mathcal{G}_{0}(-k) \, \mathcal{G}_{0}(q-k) \, T_{11}(q)
\nonumber \\
& \simeq & - 2 \sum_{k}  \mathcal{G}_{0}(k)^{2} \mathcal{G}_{0}(-k)^{2} \sum_{q} e^{i \Omega_{\nu} \eta} T_{11}(q)
\label{BEC-Popov-bosonic-self-energy-I}
\end{eqnarray}
\end{small}
\noindent
\hspace{-0.25cm} with $\eta = 0^{+}$, where the ``small-$q$'' behavior (\ref{bosonic-expression-T_11}) is picked up by the sum over $q$.
This contribution is depicted diagrammatically in Fig.~13(a).
With the help of the result (\ref{special-sums}) and introducing the definition
\begin{equation}
\sum_{q} e^{i \Omega_{\nu} \eta} \, T_{11}(q) = - \frac{8 \pi}{m^{2} a_{F}} \, n' \, ,
\label{non-condensate-density} 
\end{equation}
\noindent
of the \emph{non-condensate density} $n'$  in the broken-symmetry phase \cite{Pieri-2005}, one obtains eventually $\Sigma_{\mathrm{Popov}}^{\mathrm{B}}(0)_{11}^{(\mathrm{I})} \simeq m \, a_{F}^{2} \, n'$.
Apart from a sign difference, this result coincides with the expression obtained in Ref.~\cite{Pisani-2018} upon approaching $T_{c}$ from the normal phase, provided one replaces the total bosonic density $n/2$  
therein with the non-condensate density $n'$.

\begin{figure}[t]
\begin{center}
\includegraphics[width=8.8cm,angle=0]{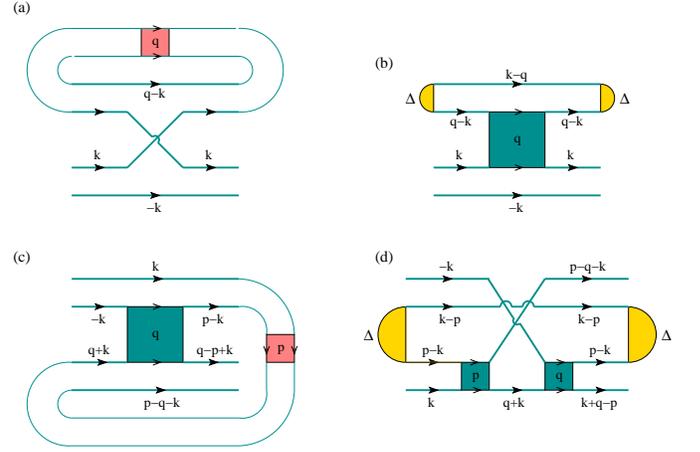}
\caption{(Color online) Graphical representation of the Popov (upper panels) and GMB (lower panels) bosonic-like self-energies in the BEC limit, which are proportional to the
                                    non-condensate density $n'$ (left panels) and to the condensate density $n_{0}$ (right panels).
                                    Here, red boxes correspond to the expression (\ref{bosonic-expression-T_11}), while blue boxes refer to the expression (\ref{asymptotic-expression-T_11})
                                    that enters the scattering processes of composite bosons.
                                    The yellow semicircles, which identify the gap parameter $\Delta$, map onto condensate lines in the case of point-like bosons.
                                    In these diagrams, we have conformed to a different convention from the rest of the paper, whereby parallel arrows of fermionic lines  
                                    signify the joint propagation of a fermion pair in the BEC limit where the system is extremely dilute.}
\label{Figure-13}
\end{center} 
\end{figure}

The second term term in Eq.~(\ref{two-terms-Popov}) becomes instead:
\begin{eqnarray}
\Sigma_{\mathrm{Popov}}^{\mathrm{B}}(\mathrm{II}) & = &
2 \left( \frac{8 \pi}{m^{2} a_{F}} \right) n_{0} \, \sum_{k,q}  \mathcal{G}_{0}(k)^{2} \, \mathcal{G}_{0}(-k) 
\nonumber \\
& \times & \mathcal{G}_{0}(q-k)^{2} \, \mathcal{G}_{0}(k-q) \, T_{11}(q)
\label{BEC-Popov-bosonic-self-energy-II}
\end{eqnarray}
\noindent
since $\Delta^{2} = \left( \frac{8 \pi}{m^{2} a_{F}} \right) n_{0}$ in the BEC limit, where $n_{0}$ is the condensate density such that $n/2 = n_{0} + n'$ \cite{Pieri-2005}.
This contribution is depicted diagrammatically in Fig.~13(b).
In this case, the ``large-$q$'' behavior (\ref{asymptotic-expression-T_11}) is picked up by the sum over $q$ in Eq.~(\ref{BEC-Popov-bosonic-self-energy-II}), and one obtains
$\Sigma_{\mathrm{Popov}}^{\mathrm{B}}(0)_{11}^{(\mathrm{II})} \simeq - 0.42 \, m \, a_{F}^{2} \, n_{0}$ according to a result given in Ref.~\cite{Pisani-2018}.

By grouping together the two contributions (\ref{BEC-Popov-bosonic-self-energy-I}) and (\ref{BEC-Popov-bosonic-self-energy-II}), we write eventually:
\begin{equation}
\Sigma_{\mathrm{Popov}}^{\mathrm{B}} = \left( \alpha^{(\mathrm{Popov})} \, n' + \beta^{(\mathrm{Popov})} \, n_{0} \right) m a_{F}^{2}
\label{BEC-Popov-bosonic-self-energy-total}
\end{equation}
\noindent
where $\alpha^{(\mathrm{Popov})} = 1$ and $\beta^{(\mathrm{Popov})} = - 0.42$.
To the extent that $n_{0} \gg n'$ for a weakly-interacting gas of composite bosons that form in the BEC limit, the term containing $n_{0}$ is dominant over that containing $n'$ as we have also verified numerically
in Fig.~\ref{Figure-5}(b) of the main text.

\vspace{0.05cm}
\begin{center}
{\bf B. GMB contribution below $T_{c}$ in the BEC limit}
\end{center}
\vspace{-0.2cm}

Analogous results can be obtained for the GMB contribution, for which it is appropriate to calculate directly the difference $ \Sigma_{\mathrm{GMB}}^{\mathrm{B}}(0)_{11} - \Sigma_{\mathrm{GMB}}^{\mathrm{B}}(0)_{12}$ given by the expression (\ref{computationally-convenient-Sigma_GMB}). 
Also in this case, two relevant contributions there arise once the expansion (\ref{approximate-BCS-propagator-11}) for $\mathcal{G}_{22}^{\mathrm{BCS}}(k)$ (and a similar one for 
$\mathcal{G}_{11}^{\mathrm{BCS}}(k)$) is considered in that expression.
We write accordingly:
\begin{equation}
\Sigma^{\mathrm{B}}_{\mathrm{GMB}} = \Sigma^{\mathrm{B}}_{\mathrm{GMB}}(\mathrm{I}) + \Sigma^{\mathrm{B}}_{\mathrm{GMB}}(\mathrm{II}) \, .
\label{two-terms-GMB}
\end{equation}
\noindent
Here, the first term coincides with that obtained for the normal phase in Ref.~\cite{Pisani-2018}, with the provision of replacing $n/2$ therein by $n'$ (plus an additional sign changes due to the different conventions we now use in the broken-symmetry phase).
One then obtains $\Sigma^{\mathrm{B}}_{\mathrm{GMB}}(\mathrm{I}) \simeq - 0.42 \, m \, a_{F}^{2} \, n'$.
This contribution is depicted diagrammatically in Fig.~13(c).

The second term in Eq.~(\ref{two-terms-GMB}), on the other hand, originates from the second term in the expansion (\ref{approximate-BCS-propagator-11}) for $\mathcal{G}_{22}^{\mathrm{BCS}}(k)$ (as well as from a similar expansion for $\mathcal{G}_{11}^{\mathrm{BCS}}(k)$), and thus contains a factor $\Delta^{2}$ proportional to $n_{0}$.
The two contributions originating in this way are depicted diagrammatically in Fig.~13(d) and are seen to coincide with each other by symmetry considerations.
The corresponding analytic expression can be obtained by a lengthly but straightforward extension of the method used in Ref.~\cite{Pisani-2018} to obtain $\Sigma^{\mathrm{B}}_{\mathrm{GMB}}$ in the normal phase.
The end result is $\Sigma^{\mathrm{B}}_{\mathrm{GMB}}(\mathrm{II}) \simeq 0.20 \, m \, a_{F}^{2} \, n_{0}$.

By grouping together the two contributions for $\Sigma^{\mathrm{B}}_{\mathrm{GMB}}$, we write in analogy to Eq.~(\ref{BEC-Popov-bosonic-self-energy-total})
\begin{equation}
\Sigma_{\mathrm{GMB}}^{\mathrm{B}} = \left( \alpha^{(\mathrm{GMB})} \, n' + \beta^{(\mathrm{GMB})} \, n_{0} \right) m a_{F}^{2} \, ,
\label{BEC-GMB-bosonic-self-energy-total}
\end{equation}
\noindent
where now $\alpha^{(\mathrm{GMB})} = - 0.42$ and $\beta^{(\mathrm{GMB})} =  0.20$.
This result, too, has been verified numerically in Fig.~\ref{Figure-5}(b) of the main text.

\vspace{0.05cm}
\begin{center}
{\bf C. Contributions to the scattering length of composite bosons below $T_{c}$}
\end{center}
\vspace{-0.2cm}

As evidenced by the way the diagrams of Fig.~13 have been drawn, the Popov and GMB results (\ref{BEC-Popov-bosonic-self-energy-total}) and (\ref{BEC-GMB-bosonic-self-energy-total}) 
can be interpreted in terms of specific scattering processes that contribute to the value of scattering length $a_{B}$ of composite bosons.
To this end, it is convenient to rewrite in the expressions (\ref{BEC-Popov-bosonic-self-energy-total}) and (\ref{BEC-GMB-bosonic-self-energy-total}):
\begin{equation}
m \, a_{F}^{2} = \left( \! \frac{m^{2} a_{F}}{8 \pi} \! \right) \, \frac{4 \pi \left(2 a_{F} \right)}{m} 
\label{rewriting}
\end{equation}
\noindent
where the factor $\left( \! \frac{m^{2} a_{F}}{8 \pi} \! \right)$ is required to comply with the structure of the pair propagator (\ref{bosonic-expression-T_11}). 
In addition, the Popov and GMB results (\ref{BEC-Popov-bosonic-self-energy-total}) and (\ref{BEC-GMB-bosonic-self-energy-total}) can be grouped together with the results (\ref{approximate-A-BEC_limit})-(\ref{special-sums}) at the mean-field level valid in the BEC limit, in such a way that the modified form of the gap equation (\ref{Popov_plus_GMB-HP}) yields for the chemical potential of composite bosons
the expression:
\begin{eqnarray}
\mu_{B} & \simeq & \frac{4 \pi (2 a_{F})}{2m} \left[ 1 + 2 \left( \beta^{(\mathrm{Popov})} + \beta^{(\mathrm{GMB})} \right) \right] \, n_{0}
\nonumber \\
& + & \frac{8 \pi (2 a_{F})}{2m} \left[ \alpha^{(\mathrm{Popov})} + \alpha^{(\mathrm{GMB})} \right] \, n'
\label{mu_B-total}
\end{eqnarray}
\noindent
where $1 + 2 \left( \beta^{(\mathrm{Popov})} + \beta^{(\mathrm{GMB})} \right) \simeq 0.56$ and $\alpha^{(\mathrm{Popov})} + \alpha^{(\mathrm{GMB})} \simeq 0.58$ in terms of the results obtained above.

From the result (\ref{mu_B-total}), we conclude that in the term proportional to $n_{0}$ pairing fluctuations beyond mean field modify the value of the scattering length for composite bosons from 
$a_{B} = 2 a_{F}$ to $a_{B} = 1.12 a_{F}$. 
In addition, they introduce a term proportional to $n'$ in which $a_{B}$ equals $1.16 a_{F}$.
On the other hand, if \emph{all} possible diagrammatic scattering processes between composite bosons were included (on top of those shown in Fig.~13), one would expect the value 
$a_{B} = 0.6 a_{F}$ to occur in both terms \cite{Kagan-2005}.

In this respect, note that in the present work the diagrams to be included in the modified gap equation have been selected in the weak-coupling limit, where the Popov and GMB bosonic self-energies were
recognized as the \emph{minimal set} needed for dressing the $t$-matrix approximation in order to recover the correct value of the gap.
It is then remarkable that the same set of diagrams recovers also the first contributions to the series of diagrams identified in Ref.~\cite{Kagan-2005}, which describes the interaction between composite bosons
in the strong-coupling limit.
Alternatively, one could have proceeded in reverse, starting from the above series of Ref.~\cite{Kagan-2005} where only diagrams with the smaller number of particle-particle propagators are retained, so as to construct with them the bosonic self-energies to be inserted in the the modified gap equation.
In this way, the Popov and GMB bosonic self-energies would have been obtained, together with a number of additional diagrams of higher-order in the small parameter $k_{F}|a_{F}|$ in the weak-coupling limit.

Note, finally, that the result (\ref{mu_B-total}) gives support to the use of the terminology Hugenholtz-Pines condition we have adopted for the modified form of the gap equation, owing to its strict analogy
with the Hugenholtz-Pines condition for point-like bosons \cite{HP-1959}. 
                                                                                                                                                                                                                                                                                                                                                                                                
\section{NUMERICAL IMPLEMENTATION OF THE POPOV AND GMB CONTRIBUTIONS TO THE GAP EQUATION}
\label{sec:appendix-C}
\vspace{-0.2cm}

In this Appendix, we cast the Popov [Eq.~(\ref{Popov-bosonic-self-energy})] and GMB [Eq.~(\ref{GMB-bosonic-self-energy_11})] expressions for the ``normal'' bosonic-like self-energy $\Sigma^{\mathrm{B}}(0)_{11}$,
as well as the GMB [Eq.~(\ref{GMB-bosonic-self-energy_12})] expression for the ``anomalous'' bosonic-like self-energy $\Sigma^{\mathrm{B}}(0)_{12}$, in a form that makes it easier to map these expressions 
for the superfluid phase with those obtained in Ref.~\cite{Pisani-2018} for the normal phase.
[As far as the GMB contribution is concerned, in practice it will be convenient to calculate directly $\Sigma_{\mathrm{GMB}}^{\mathrm{B}}(0)_{12}$ and the difference $\Sigma_{\mathrm{GMB}}^{\mathrm{B}}(0)_{11} - \Sigma_{\mathrm{GMB}}^{\mathrm{B}}(0)_{12}$, instead of calculating $\Sigma_{\mathrm{GMB}}^{\mathrm{B}}(0)_{11}$ and $\Sigma_{\mathrm{GMB}}^{\mathrm{B}}(0)_{12}$ separately.]
This mapping will speed up considerably the numerical calculation of the relevant  bosonic-like self-energies, to the extent that one can count directly on the experience nurtured with the calculation of similar quantities in the normal phase, as described in detail in Appendix A of Ref.~\cite{Pisani-2018}.

The key feature which allows this mapping to be implemented is the presence of the single-particle fermionic propagators (\ref{BCS-propagator-11}) and (\ref{BCS-propagator-12}) taken at the mean-field level 
(albeit with the replacement $\Delta^{\mathrm{BCS}} \rightarrow \Delta$) in the expressions (\ref{Popov-bosonic-self-energy}), (\ref{GMB-bosonic-self-energy_11}), and (\ref{GMB-bosonic-self-energy_12}) 
that have to be calculated.
These propagators, in turn, can be conveniently rewritten in the following form:
\begin{eqnarray}
\mathcal{G}_{11}^{\mathrm{BCS}}(k) & = & u_{\mathbf{k}}^{2} \, \tilde{\mathcal{G}}_{0}(k) - v_{\mathbf{k}}^{2} \, \tilde{\mathcal{G}}_{0}(-k)
\label{BCS-propagator-11-convenient} \\
\mathcal{G}_{12}^{\mathrm{BCS}}(k) & = & - u_{\mathbf{k}} v_{\mathbf{k}} \left( \tilde{\mathcal{G}}_{0}(k) + \tilde{\mathcal{G}}_{0}(-k) \right) \, .
\label{BCS-propagator-12-convenient} 
\end{eqnarray}
\noindent
Here, $u_{\mathbf{k}}$ and $v_{\mathbf{k}}$ are the BCS factors given by Eq.~(\ref{u-v-BCS}), while $\tilde{\mathcal{G}}_{0}$ given by Eq.~(\ref{modified-non-interacting-G}) has the same form of the
non-interacting fermionic propagator $\mathcal{G}_{0}(k) = \left( i \omega_{n} - \xi_{\mathbf{k}} \right)^{-1}$ with $\xi_{\mathbf{k}}$ replaced by $E_{\mathbf{k}}$.
In addition, the following identities hold:
\begin{eqnarray}
- \frac{1}{\Delta} \, \mathcal{G}_{12}^{\mathrm{BCS}}(k) & = & \mathcal{G}_{11}^{\mathrm{BCS}}(k) \mathcal{G}_{22}^{\mathrm{BCS}}(k) - \mathcal{G}_{12}^{\mathrm{BCS}}(k) \mathcal{G}_{12}^{\mathrm{BCS}}(k)
\nonumber \\
\mathcal{G}_{12}^{\mathrm{BCS}}(k) & = & \Delta \, \tilde{\mathcal{G}}_{0}(k) \, \tilde{\mathcal{G}}_{0}(-k) \, ,
\label{additional-BCS-identities}
\end{eqnarray}
\noindent
which can be combined together to express [cf. Eq.~(\ref{identity-BCS-non_interacting})]
\begin{equation}
\mathcal{G}_{11}^{\mathrm{BCS}}(k) \mathcal{G}_{22}^{\mathrm{BCS}}(k) - \mathcal{G}_{12}^{\mathrm{BCS}}(k) \mathcal{G}_{12}^{\mathrm{BCS}}(k) = - \tilde{\mathcal{G}}_{0}(k) \, \tilde{\mathcal{G}}_{0}(-k) \, ,
\label{useful-BCS-identity-1}
\end{equation}
\noindent
as well as to write
\begin{eqnarray}
\mathcal{G}_{11}^{\mathrm{BCS}}(k) \mathcal{G}_{22}^{\mathrm{BCS}}(k) & = & \mathcal{G}_{12}^{\mathrm{BCS}}(k)^{2} - \frac{1}{\Delta} \, \mathcal{G}_{12}^{\mathrm{BCS}}(k)
\label{useful-BCS-identity-2} \\
& = & \Delta^{2} \tilde{\mathcal{G}}_{0}(k)^{2} \, \tilde{\mathcal{G}}_{0}(-k)^{2} - \tilde{\mathcal{G}}_{0}(k) \, \tilde{\mathcal{G}}_{0}(-k) \, .
\nonumber
\end{eqnarray}
\noindent
The identity (\ref{useful-BCS-identity-1}) was already used in Section \ref{sec:G-MB-BCS-BEC} to manipulate the GMB expression for the difference $\Sigma_{\mathrm{GMB}}^{\mathrm{B}}(0)_{11} - \Sigma_{\mathrm{GMB}}^{\mathrm{B}}(0)_{12}$, in the form of the right-hand side of Eq.~(\ref{computationally-convenient-Sigma_GMB}).
Here, we rewrite that expression in an alternative form which is of better use for numerical calculations, by making the change of variables $\bar{k} = p - k$:
\begin{widetext}
\begin{equation}
\Sigma_{\mathrm{GMB}}^{\mathrm{B}}(0)_{11} - \Sigma_{\mathrm{GMB}}^{\mathrm{B}}(0)_{12}
= \sum_{\bar{k} p q} T_{11}(p) \, T_{11}(q) \, \mathcal{G}_{11}^{\mathrm{BCS}}(p+q-\bar{k}) \, \mathcal{G}_{22}^{\mathrm{BCS}}(-\bar{k})
\tilde{\mathcal{G}}_{0}(\bar{k}-q) \, \tilde{\mathcal{G}}_{0}(q-\bar{k}) \, \tilde{\mathcal{G}}_{0}(p-\bar{k}) \, \tilde{\mathcal{G}}_{0}(\bar{k}-p) \, .
\label{computationally-convenient-Sigma_GMB-final-I} 
\end{equation}
\noindent
The form (\ref{BCS-propagator-11-convenient}) of $\mathcal{G}_{11}^{\mathrm{BCS}}(k)= - \mathcal{G}_{22}^{\mathrm{BCS}}(-k)$ can further be used on the right-hand side of 
Eq.~(\ref{computationally-convenient-Sigma_GMB-final-I}), to express the integrand therein as $T_{11}(p) \, T_{11}(q)$ times a linear combination of products of six $\tilde{\mathcal{G}}_{0}$ with appropriate arguments.
In this way one ends up with the following expression:
\begin{equation}
\Sigma_{\mathrm{GMB}}^{\mathrm{B}}(0)_{11}  - \Sigma_{\mathrm{GMB}}^{\mathrm{B}}(0)_{12} = - \!\! \int \! \frac{d\mathbf{k}}{(2 \pi)^{3}}  \sum_{p,q} T_{11}(p) T_{11}(q)  
\, \mathcal{J}(E_{\mathbf{k}},E_{\mathbf{p-k}},E_{\mathbf{q-k}},E_{\mathbf{p+q-k}};\Omega_{p},\Omega_{q}) 
\label{computationally-convenient-Sigma_GMB-final-II} 
\end{equation}
\noindent
where
\begin{eqnarray}
\mathcal{J}(E_{\mathbf{k}},E_{\mathbf{p-k}},E_{\mathbf{q-k}},E_{\mathbf{p+q-k}};\Omega_{p},\Omega_{q})
& = & \left\{ u_{\mathbf{k}}^{2} u_{\mathbf{p+q-k}}^{2} \, J(E_{\mathbf{k}},E_{\mathbf{p-k}},E_{\mathbf{q-k}},E_{\mathbf{p+q-k}};\Omega_{p},\Omega_{q}) \right.   
\nonumber \\
& + & v_{\mathbf{k}}^{2} u_{\mathbf{p+q-k}}^{2} \, J(-E_{\mathbf{k}},E_{\mathbf{p-k}},E_{\mathbf{q-k}},E_{\mathbf{p+q-k}};\Omega_{p},\Omega_{q}) 
\nonumber  \\  
& + & u_{\mathbf{k}}^{2} v_{\mathbf{p+q-k}}^{2} \, J(E_{\mathbf{k}},E_{\mathbf{p-k}},E_{\mathbf{q-k}},-E_{\mathbf{p+q-k}};\Omega_{p},\Omega_{q}) 
\nonumber \\
& + & \left. v_{\mathbf{k}}^{2} v_{\mathbf{p+q-k}}^{2} \, J(-E_{\mathbf{k}},E_{\mathbf{p-k}},E_{\mathbf{q-k}},- E_{\mathbf{p+q-k}};\Omega_{p},\Omega_{q})  \right\} 
\label{notation-mathcal-J} 
\end{eqnarray}
\noindent
with the short-hand notation
\begin{equation}
J(E_{\mathbf{k}},E_{\mathbf{p-k}},E_{\mathbf{q-k}},E_{\mathbf{p+q-k}};\Omega_{p},\Omega_{q}) = T \, \sum_{\omega_{n}}
\tilde{\mathcal{G}}_{0}(p+q-k) \, \tilde{\mathcal{G}}_{0}(k) \, \tilde{\mathcal{G}}_{0}(\bar{k}-q) \, \tilde{\mathcal{G}}_{0}(q-\bar{k}) \, \tilde{\mathcal{G}}_{0}(p-\bar{k}) \, \tilde{\mathcal{G}}_{0}(\bar{k}-p) \, .
\label{notation-J}
\end{equation}
\noindent
in analogy to a similar notation introduced in Appendix A of Ref.~\cite{Pisani-2018}.

The identity (\ref{useful-BCS-identity-2}) can be used in conjunction with (\ref{additional-BCS-identities}) to manipulate the expression of $\Sigma_{\mathrm{GMB}}^{\mathrm{B}}(0)_{12}$ given by Eq.~(\ref{GMB-bosonic-self-energy_12}) 
(where we note that the two terms on the right-hand side are equal to each other owing to the symmetry of the integrand under the interchange $k \leftrightarrow k'$).
With the change of variables $k = q - \bar{k}$, $k' = p - \bar{k}$, $k'' = - \bar{k}$, we obtain:
\begin{eqnarray}
\Sigma_{\mathrm{GMB}}^{\mathrm{B}}(0)_{12} & = &
2 \sum_{\bar{k}pq} T_{11}(p) \, T_{11}(q) \, \mathcal{G}_{11}^{\mathrm{BCS}}(p+q-\bar{k}) \, \mathcal{G}_{22}^{\mathrm{BCS}}(-\bar{k})
\mathcal{G}_{11}^{\mathrm{BCS}}(q-\bar{k}) \, \mathcal{G}_{22}^{\mathrm{BCS}}(q-\bar{k}) \, \mathcal{G}_{12}^{\mathrm{BCS}}(p-\bar{k}) \, \mathcal{G}_{12}^{\mathrm{BCS}}(p-\bar{k}) 
\nonumber \\
& = & 2 \sum_{kpq} T_{11}(p) \, T_{11}(q) \, \mathcal{G}_{11}^{\mathrm{BCS}}(p+q-k) \, \mathcal{G}_{22}^{\mathrm{BCS}}(-k)
\nonumber \\
& \times & \left\{ \Delta^{4} \, \tilde{\mathcal{G}}_{0}(q-k)^{2} \, \tilde{\mathcal{G}}_{0}(k-q)^{2} - \Delta^{2} \, \tilde{\mathcal{G}}_{0}(q-k) \, \tilde{\mathcal{G}}_{0}(k-q) \right\} \, 
\tilde{\mathcal{G}}_{0}(p-k)^{2} \, \tilde{\mathcal{G}}_{0}(k-p)^{2} \, .
\label{GMB-bosonic-self-energy_12-final-I} 
\end{eqnarray}
\noindent
Here, the factor $\mathcal{G}_{11}^{\mathrm{BCS}}(p+q-k) \, \mathcal{G}_{22}^{\mathrm{BCS}}(-k)$ can be expressed in terms of products of two $\tilde{\mathcal{G}}_{0}$ using 
Eq.~(\ref{BCS-propagator-11-convenient}), 
while the squares of $\tilde{\mathcal{G}}_{0}(q-k) \, \tilde{\mathcal{G}}_{0}(k-q)$ and of $\tilde{\mathcal{G}}_{0}(p-k) \, \tilde{\mathcal{G}}_{0}(k-p)$ can be reduced to the products of two $\tilde{\mathcal{G}}_{0}$
by noting that
\begin{equation}
\left( \tilde{\mathcal{G}}_{0}(k) \, \tilde{\mathcal{G}}_{0}(-k) \right)^{2} = \left( \frac{1}{\omega_{n}^{2}+E_{\mathbf{k}}^{2}} \right)^{2}
= - \frac{\partial}{\partial E_{\mathbf{k}}^{2}} \left( \frac{1}{\omega_{n}^{2}+E_{\mathbf{k}}^{2}} \right) = 
- \frac{\partial}{\partial E_{\mathbf{k}}^{2}} \left( \tilde{\mathcal{G}}_{0}(k) \, \tilde{\mathcal{G}}_{0}(-k) \right) \, .
\label{manipulation-squares}
\end{equation}
\noindent
With the definitions (\ref{notation-mathcal-J}) and (\ref{notation-J}), the expression (\ref{GMB-bosonic-self-energy_12-final-I}) can be eventually cast in the form:
\begin{eqnarray}
\Sigma_{\mathrm{GMB}}^{\mathrm{B}}(0)_{12} & = & 2 \int \! \frac{d\mathbf{k}}{(2 \pi)^{3}} \sum_{p,q} T_{11}(p) T_{11}(q)
\left\{ \Delta^{4} \frac{\partial}{\partial E_{\mathbf{q-k}}^{2}} \frac{\partial}{\partial E_{\mathbf{p-k}}^{2}} 
\mathcal{J}(E_{\mathbf{k}},E_{\mathbf{p-k}},E_{\mathbf{q-k}},E_{\mathbf{p+q-k}};\Omega_{p},\Omega_{q}) \right.
\nonumber \\
& + & \left. \Delta^{2} \frac{\partial}{\partial E_{\mathbf{p-k}}^{2}} \mathcal{J}(E_{\mathbf{k}},E_{\mathbf{p-k}},E_{\mathbf{q-k}},E_{\mathbf{p+q-k}};\Omega_{p},\Omega_{q})  \right\} \, .
\label{GMB-bosonic-self-energy_12-final-II} 
\end{eqnarray}
\end{widetext}

The Popov bosonic-like self-energy (\ref{Popov-bosonic-self-energy}) can also be manipulated along similar lines through a repeated use of the identity (\ref{BCS-propagator-11-convenient}), in order to bring it to the form of a linear combination of the corresponding expression valid in the normal phase above $T_{c}$ as discussed in Ref.~\cite{Pisani-2018}, apart again from the replacement
$\mathcal{G}_{0} \rightarrow \tilde{\mathcal{G}}_{0}$.

The numerical calculation of the expressions (\ref{computationally-convenient-Sigma_GMB-final-II}) and (\ref{GMB-bosonic-self-energy_12-final-II}) for the GMB contribution 
(as well as of the corresponding expression for the Popov contribution) can now proceed following step by step the prescriptions given in detail in Appendix A of Ref.~\cite{Pisani-2018}, with the only provision of replacing the cutoff $k_{c}$ defined in Eq.~(A4) therein with the new value $k_{c} = \sqrt{2m \left[ \mu^{2} + \mathrm{max}(\Delta,T)^{2} \right]^{1/2} }$, to account for the presence of a finite value of $\Delta$ in the broken-symmetry phase.


\newpage


\begin{thebibliography}{99}
\bibitem{footnote-1} In the broken-symmetry phase, where particle and hole excitations are intimately related to each other, the distinction between particle-particle and particle-hole diagrams appears 
                                only conventional. For this reason, the arrows decorating the diagrams of Figs.~\ref{Figure-1} and \ref{Figure-2} have only the meaning of pointing from the 
                                second to the first argument of the fermionic propagators.
\bibitem{footnote-4} The use of single-particle propagators of the mean-field form (\ref{BCS-propagator-11}) and (\ref{BCS-propagator-12}) in the expressions (\ref{Sigma_11-t_matrix}) for the $t$-matrix, 
                                 (\ref{Popov-approximate-propagators}) for the Popov contribution, and (\ref{fermionic-self-energy-GMB})-(\ref{approximate-G_12}) for the GMB contribution (with 
                                 $\Delta^{\mathrm{BCS}}$ replaced by $\Delta$ in each case), entails a partial re-summation of diagrams (although not pursued in a fully self-consistent way).
\bibitem{footnote-5} A different sign convention is here used for $\Sigma^{\mathrm{B}}$ with respect to that of Ref.~\cite{Pisani-2018} for the corresponding bosonic-like self-energy 
                                 in the normal phase.
\bibitem{footnote-6} We have verified that the values of $T_{c}$, obtained here in the broken-symmetry phase by letting $\Delta \rightarrow 0$ in the modified form of the gap equation, coincide with those obtained in 
                                 Ref.~\cite{Pisani-2018} when $T_{c}$ is approached from the normal phase.
\end{thebibliography}
\end{document}